\documentclass[review]{elsarticle}
\usepackage{lineno,hyperref}
\usepackage{graphicx}
\usepackage{subfig}
\usepackage[linesnumbered,ruled,lined]{algorithm2e}
\usepackage{float}
\usepackage{tikz}
\usepackage{epstopdf}
\usepackage{amssymb}
\usepackage{amsmath}
\usetikzlibrary{arrows.meta}
\journal{Comput. Methods Appl. Mech. Eng.}
\newcommand{\tr}{\mathop{\mathrm{tr}}}

\newcommand{\sign}{\mathop{\mathrm{sign}}}
\newcommand{\dist}{\mathop{\mathrm{dist}}}
\newcommand{\rand}{\mathop{\mathrm{random}}}
\begin{document}
	\begin{frontmatter}
		\title{A multi-order smoothed particle hydrodynamics method for cardiac electromechanics with the Purkinje network}
		\author[myfirstaddress]{Chi Zhang }
		\ead{c.zhang@tum.de}
		\author[mysecondaryaddress]{Hao Gao}
		\ead{hao.gao@glasgow.ac.uk}
		\author[myfirstaddress]{Xiangyu Hu \corref{mycorrespondingauthor}}
		\ead{xiangyu.hu@tum.de}
		\address[myfirstaddress]{Department of Mechanical Engineering, 
			Technical University of Munich, 85748 Garching, Germany}
		\address[mysecondaryaddress]{School of Mathematics \& Statics, 
		University of Glasgow, Glasgow, UK}
		\cortext[mycorrespondingauthor]{Corresponding author.}
		\begin{abstract}			
			In previous work, 
			Zhang et al. (2021) \cite{zhang2021integrative} developed an integrated smoothed particle hydrodynamics (SPH) method 
			to address the simulation of the principle aspects of cardiac function, 
			including electrophysiology, passive and active mechanical response of the myocardium.  
			As the inclusion of the Purkinje network in electrocardiology is recognized as fundamental 
			to accurately describing the electrical activation in the right and left ventricles, 
			in this paper, 
			we present a multi-order SPH method to handle the electrical propagation through the Purkinje system and in the myocardium 
			with monodomain/monodomain coupling strategy. 
			We first propose an efficient algorithm for network generation on arbitrarily complex surface 
			by exploiting level-set geometry representation and cell-linked list neighbor search algorithm. 
			Then, 
			a reduced-order SPH method is developed to solve the one-dimensional monodomain equation to 
			characterize the fast electrical activation through the Purkinje network. 
			Finally, 
			a multi-order coupling paradigm is introduced to capture the coupled nature of potential propagation 
			arising from the interaction between the network system and the myocardium.
			A set of numerical examples are studied to assess the computational performance, accuracy and versatility of the proposed methods. 
			In particular, 
			numerical study performed in realistic left ventricle demonstrates that the present method features all the physiological issues that characterize a heartbeat simulation,
			including the initiation of the signal in the Purkinje network and the systolic and diastolic phases.
			As expected, the results underlie the importance of using physiologically realistic Purkinje system for modeling cardiac functions.
		\end{abstract}
	\end{frontmatter}
%
%
\section{Introduction}\label{sec:introduction}
Cardiac diseases due to complex mechanisms represent 
one of the most important category of problems in public health, 
effecting millions of people each year according to 
the reports of World Health Organization (WHO) \cite{who}. 
Computational study of cardiac function has received tremendous efforts
and is recognized as the community's \textit{next microscope, only better} \cite{trayanova2011whole}. 
Since the heart's physiology involves multiple physics systems, 
e.g. 
electrophysiology, 
(passive and active) mechanics and hemodynamics,
an effective integrated computational model is very challenging 
and requires accurate coupling of all these biophysical systems 
and asks for advanced numerical techniques \cite{quarteroni2017cardiovascular}. 
Despite of substantial efforts on integrated cardiac modeling, 
including fluid-structure interaction (FSI) and fluid-structure-electrophysiology interaction (FSEI), 
by applying the finite-element method (FEM) \cite{gao2017coupled, santiago2018fully} 
and the immersed-boundary method (IBM) \cite{viola2020fluid}. 
An integrative model capable of simulating the fully coupled cardiac function 
is still in its infancy due to the meshing bottlenecks of the FEM 
and the Lagrangian-Eulerian mismatches on the kinematics of the IBM. 

As an alternative, 
the meshless, fully Lagrangian smoothed particle hydrodynamics (SPH) \cite{lucy1977numerical, gingold1977smoothed} method 
has shown peculiar advantages in handling multi-physics problems \cite{liu2019smoothed, zhang2021sphinxsys} 
thanks to its very feature of representing each sub-system by an ensemble of particles. 
Since its original inception by Lucy \cite{lucy1977numerical} 
and Gingold and Monaghan \cite{gingold1977smoothed} for astrophysical applications,  
the SPH method has been successfully applied in a broad variety of applications 
ranging from fluid mechanics \cite{monaghan1994simulating, hu2006multi, zhang2017weakly, zhang2019weakly} 
and solid dynamics \cite{monaghan1992smoothed, randles1996smoothed, bonet2000correction, benz1995simulations, zhang2017generalized}
to multi-phase flows \cite{rezavand2020weakly, peng2021particle, zheng2021novel} 
and FSI \cite{antoci2007numerical, toma2021fluid, zhang2020efficient, zhang2021improved}.
More recently, 
Zhang et al. \cite{zhang2021integrative} developed an integrative SPH method for cardiac function 
and demonstrated its robust and accuracy in dealing with the following aspects : 
(i) correct capturing of the stiff dynamics of the transmembrane potential and the gating variables, 
(ii) robust predicting of the large deformations and the strongly anisotropic behavior of the myocardium, 
(iii) proper coupling of the electrical excitation and the tissue mechanics for electromechanical feedback. 
This achievement render the SPH method a potential and powerful alternative that can augment the current line of total heart modeling. 
As the inclusion of the Purkinje network in modeling of the cardiac electrophysiology has been recognized as fundamental 
to accurately describing the electrical activation in the left and right ventricles \cite{clayton2011models, abboud1991simulation, vergara2014patient}, 
developing proper SPH model to handle the electrical propagation through the Purkinje network and in the myocardium 
is essential for developing an meshless total heart simulator and exploring possible clinical applications. 

The electrical activation in the human heart originates in the sinoatrial (SA) node located in the right atrium, 
travels through the atria and enters the atrioventricular (AV) node \cite{tawara1906reizleitungssystem}. 
Through the His bundle, 
the AV node is connected to the Purkinje network which branches from the basal septum into the left and right ventricles. 
In physiological condition, 
the action potential travels along the Purkinje network and enters the ventricular muscle through the Purkinje-muscle junctions (PMJ) \cite{clayton2011models}. 
The Purkinje network is located in the sub-endocardium and composed of specialized fast-conducting cells to conduct the potential wave 
efficiently and rapidly \cite{hall1971rapid}. 
As a key component of the cardiac excitation system, 
the Purkinje network plays a key role in both physiological excitation and life threatening pathological excitation, 
i.e., arrhythmia \cite{quarteroni2017cardiovascular}. 
Therefore, 
developing algorithms for network generation on complex endocardial surface 
and capturing the coupled nature of potential propagation arising from the interaction 
between the Purkinje network and the myocardium is a fundamental task for realistic cardiac modeling. 

Concerning the network generation, 
several algorithms have been developed in the past decades since the first observation century ago \cite{tawara1906reizleitungssystem} 
of the Purkinje network for both visualization and simulation purposes \cite{ccetingul2011estimation}.  
More specifically, 
three approaches have been proposed for the Purkinje network generation, 
namely, 
patient-specific segmentation from ex-vivo images \cite{bordas2011rabbit, early2001image}, 
manual procedure based on anatomical knowledge \cite{ten2008modelling, berenfeld1998purkinje} and 
computational algorithms based on the fractal law \cite{lindenmayer1968mathematical, ijiri2008procedural, costabal2016generating,sebastian2011construction, palamara2015effective}. 
Among these approaches, 
only the first one allows to recover patient-specific information, 
however, 
it is limited by the fact that there is no in vivo image technique available to fully reconstruct the Purkinje network's structure \cite{palamara2015effective}. 
Having the inherent complexity of the Purkinje network in mind, 
the manual procedure is excessively complicated and time consuming \cite{costabal2016generating}. 
Notwithstanding the difficulty to recover patient-specific observation, 
the fractal law based algorithm, 
e.g. fractal tree \cite{costabal2016generating}, 
has received more and more attention due to its easy implementation and versatility of 
incorporating with available code for computational cardiac electrophysiology \cite{costabal2016generating, palamara2014computational}.
The fractal tree algorithm was first adapted by Abboud et al. \cite{abboud1991simulation} 
to create the Purkinje network to study high-frequency electrocardiograms. 
In their work \cite{abboud1991simulation}, 
the network consists of straight segments and is placed in a simplified ventricle. 
Then, 
Ijiri et al. \cite{ijiri2008procedural} introduced non-straight branches with controllable curvature 
to generate hierarchical network. 
One notable improvement contributed to Palamara et al. \cite{palamara2015effective} 
is on creating patient-specific Purkinje network \cite{vergara2014patient} 
by using clinical measurements of the electrical activation in the ventricle to locate the PKJs.
However, 
all these algorithms can only generate network for simplified ventricles with regular and smooth surface. 
More recently, 
Costabal et al. \cite{costabal2016generating} extended the fractal tree algorithm to create network on irregular surface 
by introducing a second-step projection. 
However, this algorithm is excessively computational expensive due to the time consuming triangle search 
for projecting each newly created node. 

As regards the cardiac modeling with inclusion of the Purkinje network, 
numerical studies have been mainly focused on the myocardium electrosphysiology with different coupling strategies, 
e.g. 
the eikonal/eikonal model \cite{vergara2014patient}, 
the eikonal/monodomain model \cite{sebastian2011construction}, 
the bidomain/bidomain model \cite{bordas2012bidomain}, 
the monodomain/bidomain model \cite{vigmond2007construction}, 
and the monodomain/monodomain model \cite{romero2010effects, vergara2016coupled}. 
Here, the first model refers to the one used for the Purkinje network and the second to that applied for the myocardium. 
As reported by Vergara et al. \cite{vergara2016coupled}, 
the monodomain/monodomain coupling model, 
termed as MM model hereafter, 
is able to capture many characteristic features of the electrical propagation in the ventricles, 
and allows to highly reduce the computational time with respect to the eikonal and bidomain models.
Another notable feature of the MM model is that it is particularly suited in view of the electromechanical coupling 
which is one of the main object of this work. 
Therefore, 
we consider the MM model in this paper. 
Notwithstanding significant efforts have been devoted to the numerical studies of cardiac electrosphysiology with inclusion of the Purkinje network, 
rare works in the literature have been devoted to study the Purkinje network's effects on 
the mechanical contraction of the ventricles.  
Usyk et al. \cite{usyk2002computational} included the fast conduction of the Purkinje network in a numerical model of cardiac electromechanics 
through a surrogate spatial modification of the myocardial conduction property. 
More recently, 
Landjuela et al. \cite{landajuela2018numerical} 
conducted a numerical study of the electromechanical coupling in the left ventricle with presence of the Purkinje network 
by applying the monodomain/bidomain model within the FEM framework.  
Despite relentless progress in computational cardiac electrophysiology and electromechanics 
with inclusion of the Purkinje network within the FEM framework, 
there is no SPH model has been developed, 
to the best knowledge of the authors, 
to study the electrical activation through the Purkinje network and in the myocardium 
despite the fact that the SPH method has been recognized as 
an emerging and promising alternative for cardiac modeling \cite{zhang2021integrative, mao2016fluid, lluch2019breaking, zhang2019meshfree, lluch2020calibration} 
and other biomechanics applications \cite{zhang2019meshfree}.

In this work, 
we start from developing an efficient algorithm based on the fractal law for network generation on non-smooth surface 
by adapting level-set methods and exploiting cell-link list (CLL) scheme. 
With the geometry representation using level-set method, 
a three-dimensional fractal tree can be projected onto arbitrarily complex surface, 
allowing the Purkinje network generation on the endocardial surface of realistic ventricles. 
Another key feature of exploiting level-set method is that the time-consuming neighboring triangles search for each newly 
created node for second-step node projection \cite{costabal2016generating} is avoided. 
Different with Ref. \cite{costabal2016generating} where a k-d tree scheme is applied for nearest node search, 
the CLL scheme is adapted in the present algorithm to incorporate with the SPH framework. 
Subsequently, 
we introduce a reduced-order SPH method for solving one-dimensional monodomain equation on linear structure in three-dimensional space. 
The key idea is to constrain the degree of freedom along one space dimension other than applying the one-dimensional discretization. 
With exploiting the operator splitting combined with reaction-by-reaction splitting and the anisotropic diffusion SPH discretization proposed in Ref. \cite{zhang2021integrative}, 
the present reduced-order SPH method can correctly capture the fast electrical activation in the Purkinje network. 
Furthermore, 
a multi-order coupling scheme is developed for MM coupling in the network and myocardium interactions. 
More precisely, 
the terminal particles of the reduced-order SPH model of the network, 
which represent the PKJs, 
take the roles as excitation sources of current flux to the myocardium particles. 
To optimize the computational efficiency, 
a multi-time stepping scheme is proposed for time integration of the corresponding electrophysiology and electromechanics coupling problems.
Ultimately, 
the proposed multi-order SPH method is integrated to predict the active response of myocardium 
by implementing the active stress approach \cite{zhang2021integrative, nash2004electromechanical}. 
A set of numerical examples, 
e.g. 
the potential propagation in a myocardium fiber, 
cubiod myocardium with inclusion of a generic network, 
electrophysiology and electromechanics in a realist left ventricle with inclusion of the Purkinje network 
are computed to demonstrate the accuracy, robustness and feasibility of the proposed multi-order SPH method. 

This manuscript is organized as follows. 
Section \ref{sec:governingeq} introduces the basic principles of the kinetics 
and the governing equations describing the evolution of the transmembrane potential,  
and the active and pass mechanical responses of the tissue. 
Section \ref{sec:network} presents the efficient algorithm for network generation. 
In Section \ref{sec:reducedoder}, 
the proposed reduced-order SPH method for solving one-dimensional monodomainq equation is fully described. 
Then, 
the multi-order coupling algorithm and multi-time stepping scheme are detailed in Section \ref{sec:multiorder}. 
A set of examples are included in Section \ref{sec:examples} and 
the concluding remarks and a summary of the key contributions of this paper are given in Section \ref{sec:conclusion}. 
For a better comparison and future openings for in-depth studies, 
all the computational codes and data-sets accompanying this work are released 
in the repository of SPHinXsys \cite{zhang2021sphinxsys, zhang2020sphinxsys} at \url{https://www.sphinxsys.org}.
%
%
\section{Governing equations}\label{sec:governingeq}
In this section, 
we briefly summarize the governing equations for the passive and active mechanical response of the myocardium, 
and the electrical activations through the Purkinje network and in the myocardium. 
\subsection{Kinematics}\label{sec:kinematics}
To characterize the deformation of a continuum, 
a material point's initial position
$\mathbf r^0$ is defined in the initial reference configuration, 
and its current position  $\mathbf r$ in the deformed configurations.
Note that the superscript $\left( \bullet \right)^0$ denotes the quantities at the reference state hereafter. 
Then the deformation tensor $\mathbb F$ can be defined by the gradient of current position with respect to the initial reference configuration as 
\begin{equation} \label{eq:deformationtensor}
	\mathbb F = \frac{\partial \mathbf r}{\partial \mathbf r^0} = \nabla^0 \mathbf u  + \mathbb I,
\end{equation}
where $\nabla^0$ denotes the spatial gradient operator with respect to the initial reference configuration, 
$\mathbf u = \mathbf r - \mathbf r^0$ the displacement of the material point and $\mathbb I$ the unit matrix. 
Having the deformation tensor $\mathbb F$, 
the left Cauchy-Green deformation tensor is given by 
\begin{equation} \label{eq:cauchy-green}
	\mathbb C = \mathbb F^T \cdot \mathbb F. 
\end{equation} 
Associated with $\mathbb C$, 
there are the principle invariants, i.e., 
\begin{equation}\label{principle-invariants}
	I_1 = \tr \mathbb C, \quad I_2 = \frac{1}{2}\left[I^2_1 - \tr(\mathbb C^2)\right], \quad I_3 = \det(\mathbb C) = J^2,
\end{equation}
where $J = \det(\mathbb F)$, 
and 3 other independent invariants due to the directional preferences
\begin{equation}\label{extra-principle-invariants}
		I_{ff}  =   \mathbb C : \mathbf f^0 \otimes\mathbf f^0,
		\quad I_{ss}  =   \mathbb C : \mathbf s^0 \otimes\mathbf s^0,
		\quad I_{fs}  =   \mathbb C : \mathbf f^0 \otimes\mathbf s^0, 
\end{equation}
where $\mathbf f^0$ and $\mathbf s^0$  are the undeformed myocardial fiber and sheet unit direction, respectively. 
\subsection{Electromechanics}\label{sec:electromechanics}
In the total Lagrangian framework, 
the momentum conservation equation for mechanical response of the myocardium can be expressed as
\begin{equation}\label{eq:mechanical-mom}
	\frac{\text d \mathbf v}{\text d t}  =  \frac{1}{\rho^0 } \nabla^0 \cdot \mathbb P^T ,
\end{equation}
where $\frac{\text d \mathbf v}{\text d t}$ is the material derivative, 
$\rho^0$ the density at the initial state and $\mathbb{P}$ the first Piola-Kirchhoff stress tensor. 

To characterize the active mechanical response of the myocardium, 
we consider the active stress approach proposed by Nash and Panfilov \cite{nash2004electromechanical} 
where the first Piola-Kirchhoff stress $\mathbb{P}$ can be decomposed into passive and active parts as
\begin{equation}
	\mathbb P = \mathbb P_p + \mathbb P_a .
\end{equation}
Here, 
the passive component $\mathbb P_p$ describes the stress required to obtain a given deformation of the passive response, 
and the active component  $\mathbb P_a$ denotes the tension induced by the electrical activation. 
	
To model the passive mechanical response,    
we modify the strain energy function of Holzapfel and Odgen\cite{holzapfel2009constitutive} 
by ensuring the stress vanishes in the reference configuration and encompassing the finite extensibility \cite{zhang2021integrative} as 
\begin{align}\label{eq:new-muscle-energy}
	\mathfrak W = &  \frac{a}{2b}\exp\left[b (I_1 - 3 )\right] - a \ln J  + \frac{\lambda}{2}(\ln J)^{2} \nonumber + \\
	& \sum_{i = f,s} \frac{a_i}{2b_i}\{\text{exp}\left[b_i\left(\mathit{I}_{ii}- 1 \right)^2\right] - 1\} \nonumber + \\
	& \frac{a_{fs}}{2b_{fs}}\{\text{exp}\left[b_{fs}\mathit{I}^2_{fs} \right] - 1\} ,
\end{align}
where $\lambda$ is the Lamé parameter. 
Also, 
$a$, $b$, $a_f$, $b_f$, $a_s$, $b_s$, $a_{fs}$ and $b_{fs}$ are proper positive material constants 
with the $a$ parameters having dimension of stress and the $b$ parameters being dimensionless. 	
Then, 
it is easy to derive the second Piola-Kirchhoff stress $\mathbb S$ as 
\begin{equation}\label{eq:second-PK}
	\mathbb S = 2\sum_j \frac{\partial \mathfrak W}{\partial \mathit I_j} \frac{\partial \mathit I_j}{\partial \mathbb C} -p\mathbb C^{-1}; 
	\quad j = \left\lbrace 1, ff,ss,fs\right\rbrace ; 
	\quad  p = \frac{\partial \mathfrak W}{\partial J}.
\end{equation}
Subsequently, 
the passive first Piola-Kirchhoff stress $\mathbb P_p$ is defined as
\begin{equation}\label{eq:passive}
			\mathbb P_p = \mathbb F \mathbb S.
\end{equation} 

Following the active stress approach \cite{nash2004electromechanical}, 
the active first Piola-Kirchhoff stress $\mathbb P_a$ is defined as
\begin{equation}\label{eq:active}
	\mathbb P_a = T_a \mathbb F \mathbf f^0 \otimes \mathbf f^0,
\end{equation}
where $T_a$ represents the active cardiomyocite contraction stress 
and its evolution is governed by an ordinary differential equation (ODE) defined as 
\begin{equation}
	\frac{\text d T_a}{\text d t} = \epsilon\left(V_m\right)\left[k_a\left(V_m - V_r \right) - T_a\right].
\end{equation}
Here, 
the parameters $k_a$ and ${V}_r$ control the maximum active force and the resting transmembrane  potential. 
Note that the activation function is given by \cite{wong2011computational}
\begin{equation} \label{eq:epsilon}
	\epsilon\left(V_m \right) = \epsilon_0 + \left(\epsilon_{\infty} -\epsilon_{-\infty} \right) \exp \{-\exp \left[-\xi\left(V_m - \overline{V}_m\right)\right]\} ,
\end{equation} 
where the limiting values $\epsilon_{-\infty}$ at $V_m \rightarrow -\infty$ and $\epsilon_{\infty}$ at $V_m \rightarrow \infty$, 
the phase shift $\overline{V}_m$ and the transition slope $\xi$ will ensure a smooth activation of the muscle traction. 
\subsection{Monodomain equation}\label{sec:monodomain}
Following the work of Vergara et al. \cite{vergara2016coupled}, 
we consider MM model to characterize the electrical activation through the Purkinje network and in the myocardium. 
In monodomain equation, 
the evolution of the transmembrane potential $V_m$ is governed by a coupled system of partial differential equations (PDEs) written as
\begin{equation}\label{eq:monodomain}
	C_m \frac{\text d V_m}{\text d t} = \nabla^0 \cdot  \left( \mathbb{D}  \nabla^0 V_m\right)  + I_{ion},
\end{equation}
where $C_m$ denotes the capacitance of the cell membrane, 
$\mathbb{D} $ the conductivity coefficient and 
and $I_{ion}$ the ionic current. 
For the electrical activation through the Purkinje network, 
the conductivity coefficient is $\mathbb{D} = d^P_{iso} \mathbb{I}$ 
implying a one-dimensional conductivity. 
For the electrical activation in the myocardium, 
the conductivity coefficient is defined with respect to the initial reference configuration 
by $\mathbb{D} = d^M_{iso} \mathbb{I} + d^M_{ani} \mathbf{f}^0 \otimes \mathbf{f}^0$ 
with $ d^M_{iso}$ denoting the isotropic contribution 
and $d^M_{ani}$ the anisotropic contribution to account for conductivity along fiber direction $\mathbf{f}^0$. 

To close the system of Eq. \eqref{eq:monodomain}, 
we apply the Aliev-Panfilow model \cite{aliev1996simple} 
which has been successfully implemented in the computational cardiac electrophysiology in realistic heart geometry \cite{panfilov1999three} 
and reads 
\begin{equation}\label{eq:a-p}
		\begin{cases}
			I_{ion}(V_m, w) = -k V_m(V_m - a)(V_m - 1) - w V_m \\
			\frac{\text d w}{\text d t} = g(V_m, w) = \epsilon(V_m, w)(-k V_m(V_m - b - 1) - w)
		\end{cases},
\end{equation}
where $\epsilon(V_m, w) = \epsilon_0 + \mu_1 w / (\mu_2 + V_m)$ and $k$, $a $, $b$,  $\epsilon_0$, $\mu_1$ and $\mu_2$ 
are proper constant parameters to be specified later. 
%
%
\section{Efficient algorithm for network generation}\label{sec:network}
In this section, 
an efficient algorithm based on the fractal law for network generation on arbitrarily complex surface is presented. 
We first summarize the level-set concept and its construction by parsing a computer-aided-design (CAD) model
to handle the arbitrarily complex surface representation. 
Then, 
the detailed algorithm for network generation is described. 
Finally, 
the CLL scheme for nearest node search is presented.
\subsection{Surface representation}\label{sec:levelset}
To represent the arbitrarily complex surface, 
we consider the level-set method where the geometry surface can be implicit defined 
by the zero level-set of the signed-distance function 
\begin{equation}
	\Gamma = \left\lbrace  \left(x, y, z \right)| \phi \left(x, y, z, t \right) = 0 \right\rbrace  ,
	\label{level-set}
\end{equation}
where $\phi$ is the level-set value. 
Then, 
the normal direction $\mathbf N= \left(  n_x, n_y, n_z \right) ^T$ of the surface can be computed from 
\begin{equation} \label{eq:levelset-norm}
	\mathbf N= \frac{\nabla \phi}{\left| \nabla \phi \right| }.
\end{equation}
To discretize the level-set function, 
a Cartesian background mesh is generated in the whole computational domain and 
the level-set value $\phi$ is equal to the distance from the cell center to the surface. 
It is worth noting that a negative phase with $\phi < 0$ is defined if the cell center is inside the geometry, 
while positive phase with $\phi > 0$ otherwise. 

To construct the level-set field, 
we choose to parse the polygon mesh, 
which can be in STL or OBJ format, 
of anatomical heart model following our previous work \cite{zhu2021cad, zhang2020efficient}, 
By parsing the corresponding polygon mesh, 
the distance from a cell center to the surface can be computed by iterating all triangles to find the nearest triangle 
and then find the closest point located on the triangle. 
Also, 
locating whether a cell is inside the geometry can be conducted 
by checking the sign of the dot product between the nearest triangle's norm and 
the vector pointing from the closest point located on the triangle to the given point. 
The detailed algorithm is summarized as Algorithm \ref{alg:levelset}.  
Also note that one can build its own triangulation parser or apply a proper open-source library, 
for example Simbody library \cite{sherman2011simbody}.
\begin{algorithm}[htb!]
	Build a Cartesian background mesh with proper resolution \;
	Read and parse the polygon mesh \;
	\For{each cell $i$ }
	{
		Get the cell center $\mathbf r_i$ \;
		Find the nearest triangle and its norm $\mathbf n_t$ \;
		Find the closest point $\mathbf r_t$ on the nearest triangle \;
		Set $\phi_i = \sign \left(\mathbf r_i \cdot \mathbf n_t\right) \left| \mathbf r_i - \mathbf r_t \right| $ \;
		Comput the normal direction with Eq. \eqref{eq:levelset-norm} \;
	}
	\caption{Level-set initialization by paring polygon mesh.}
	\label{alg:levelset}
\end{algorithm}
%
\subsection{Efficient algorithm for network generation}\label{sec:branch}
Based on the fractal law \cite{ijiri2008procedural, costabal2016generating},
a three-dimensional network on a non-smooth surface can be generated by iteratively generating branch,
which is represented as polylines of segments and nodes, 
with proper constraint and immediately projecting every newly created nodes onto the corresponding surface, 
e.g. endocardial surface of ventricles. 

\begin{figure}[htb!]
	\centering
	\includegraphics[trim = 4cm 7cm 1cm 4.5cm, clip, width=.95\textwidth]{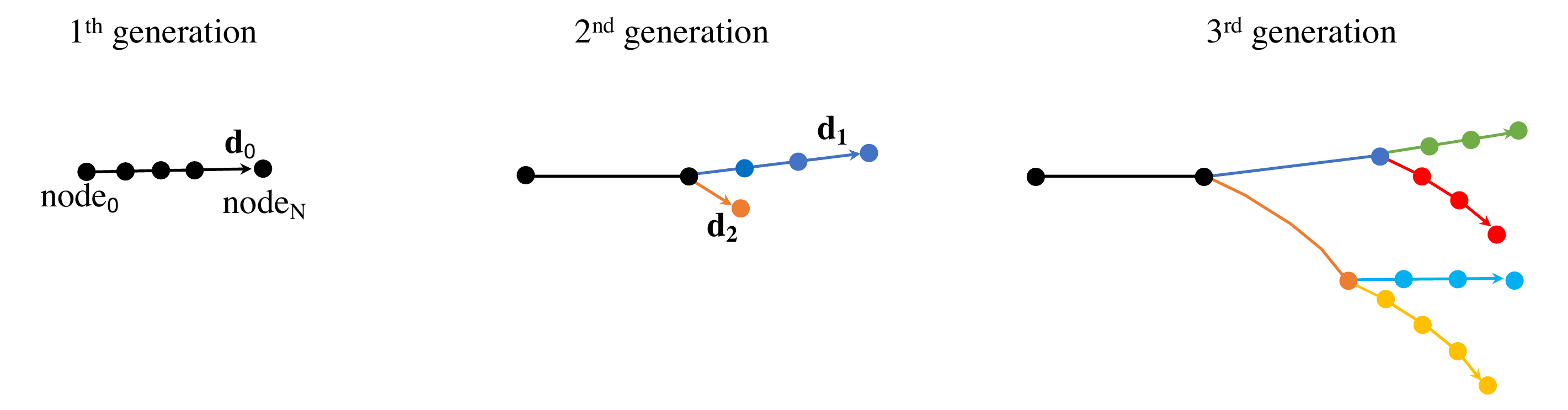}
	\caption{Schematic diagram for iteratively generation branches.}
	\label{figs:branch}
\end{figure}

With assumption of each branch consists of $N-1$ segments and $N$ nodes, 
the first branch is created by iteratively generating $N-1$ segments with a given initial node denoted as ${node}_0$, 
a prescribed growth direction $\mathbf d_0$ and a
proper segment length $l_{seg}$. 
From the terminal of the initial branch, 
i.e., ${node}_{N}$,
two \textit{child} branches in different growing direction will be created and this process will be iteratively continued on condition that 
the branch is marked as \textit{branch-to-grow}. 
Figure \ref{figs:branch} illustrates the iterative process for generating a simple network. 
Before moving to the detailed algorithm for branch growth, 
a criterion is introduced for determining whether a branch is continue to grow, 
namely marked as \textit{branch-to-grow}, or terminated. 
More precisely, 
a potential collision detection will be conducted for each newly created node, 
and the node will be removed and the branch is terminated if a collision detected. 
Otherwise, 
a branch is marked as \textit{branch-to-grow} if no collision is detected after all $N$ nodes have been fully generated. 
The collision criterion is defined as 
\begin{equation}\label{eq:collision-criterion}
	\left| \mathbf r_{new} - \mathbf r_{nearest} \right| \leq \sigma ,
\end{equation}
where $\sigma = 5  l_{seg}$ is the threshold, 
$\mathbf r_{new}$ the position of the newly created node 
and $\mathbf r_{nearest}$ the position of the nearest node of all the existing nodes in the network except the ones belonging  
to the \textit{mother} and \textit{brother} branches. 
The algorithm for the nearest node search will be presented in the following Section \ref{sec:nodesearch} by using CLL scheme to 
incorporate within the SPH framework. 

\begin{figure}[htb!]
	\centering
	\includegraphics[trim = 4cm 7cm 4cm 6cm, clip, width=.95\textwidth]{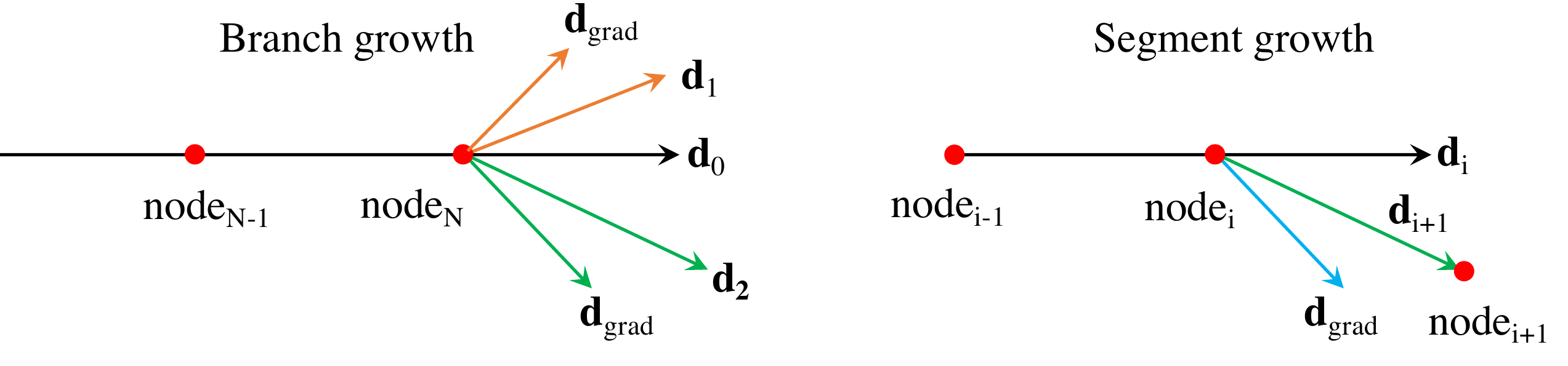}
	\caption{Schematic diagram for one \textit{branch-to-grow} branch grows into two \textit{child} branches and 
		segment growth inside a branch.
		Each \textit{branch-to-grow} branch will grow into two \textit{child} branches whose directions are given in Eq. \eqref{eq:branchgrowing}. 
		For the segment growth, 
		the growth direction is determined by the previous one and the gradient defined in Eq. \eqref{eq:graddistance}. 
		(For interpretation of the references to color in this figure legend, the reader is referred to the web version of this article.)
	}
	\label{figs:branchgrow}
\end{figure}

For all the \text{bracn-to-grow} branches, 
two \textit{child} branches will be generated at the terminal node, namely ${node}_N$, 
as shown in Figure \ref{figs:branchgrow}, 
with the growing direction defined as
\begin{equation}\label{eq:branchgrowing}
	\mathbf d_i = \frac{\hat{\mathbf d_0} + w \mathbf d_i^{grad}}{\left| \hat{\mathbf d_0} + w \mathbf d_i^{grad}\right| }; \quad i = \left\lbrace 1, 2\right\rbrace,
\end{equation}
with 
\begin{equation}\label{eq:motherdirection}
	\hat{\mathbf d_0} = \mathbf d_0 \cos\alpha + \left( \mathbf d_0 \times \mathbf N_0 \right) \sin\alpha .
\end{equation}
Here, 
$\mathbf d_0$ and $\mathbf N_0$ denote the growing direction and normal of ${node}_N$ belonging to the \textit{mother} branch, 
$i$ presents the \textit{child} branch index and $\alpha$ the growing angle. 
Also, 
$w$ the weight factor and $\mathbf d^{grad}$ the gradient of the distance are introduced 
as repulsion factors to regulates the branch curvature \cite{costabal2016generating}.  
It can be noted that the initial growing direction of a new $child$ branch is determined by 
the growing direction of ${node}_N$ belonging to its \textit{mother} branch and the repulsion factors. 
The same principle governs the segment growth, 
where the growing direction is determined by the previous one and the repulsion factors 
as shown in Figure \ref{figs:branchgrow}. 
Following Ref. \cite{costabal2016generating}, 
the gradient of the distance is computed with a central finite difference approximation
\begin{equation}\label{eq:graddistance}
	\mathbf d^{grad} = \frac{1}{2\epsilon} \left\lbrace \dist\left( \mathbf r_0 + \epsilon \mathbf e \right) - \dist\left(\mathbf r_0 - \epsilon \mathbf e\right)  \right\rbrace, 
\end{equation}
where $\epsilon = l_{seg}$ is the parameter, 
$\mathbf e$ the Cartesian basis vector, 
$\mathbf r_0$ the position of the previous node and 
$\dist\left( \bullet \right) $ function returns the distance of any given point to the closest node found in the network. 

To efficiently generate a three-dimensional network on arbitrary complex surface, 
we immediately project every newly created node onto the surface represented by the level-set with
\begin{equation}\label{eq:nodeproject}
	\begin{cases}
		\mathbf r_{project}  =  \mathbf r_{node} - \phi _{node} \mathbf N_{node}  \\
		\phi _{node} = \phi( \mathbf r_{node}) \\
		\mathbf N_{node} = \mathbf N( \mathbf r_{node})
	\end{cases},
\end{equation}
where $\phi _{node}$ and $\mathbf N_{node}$ can be interpolated from the level-set filed. 
Note that Refs. \cite{ijiri2008procedural, costabal2016generating} projected the node onto the surface by using 
the normal or vortex normal of the neighboring triangular elements. 
In this case, 
an immediate difficult relates the computational cost of searching the neighboring triangular elements. 
This time consuming process is removed in the present algorithm by exploiting the level-set method where
projection is straightforward by using trilinear data interpolation. 

Similarly to Ref. \cite{costabal2016generating}, 
three kinematic factor, 
i.e.,  
the segment number $N-1$ and length $l_{seg}$, 
the branch angle $\alpha$ and the repulsion factors $w$ and $\mathbf d^{grad} $, 
determine the shape of the network. 
To embed the network generation with the SPH framework, 
the segment number and length will be set as constant parameters without special specification. 
To introduce randomness to the network generation, 
we calculate the branch angle by adding a small random number. 
Also, 
we shuffle the order of growing branches in each generation to randomly distribute the influence of curvature with 
respect to existing nodes. 
The detailed algorithm for generating network is described in Algorithm \ref{alg:network}.
\begin{algorithm}[htb!]
	Parameter setup, 
	e.g. number of iteration $N_{iteration}$, number of segments $N$ and length of segment $l_s$ \;
	Given an initial node $\mathbf r_0$, initial direction $\mathbf d_0$ and branch angle $\alpha_0$ \;
	Do the first  generation as shown in Figure \ref{figs:branch} \;
	Add the first branch to \textit{branch-to-grow} list \;
	\For{$i \leq N_{iteration}$  }
	{
		Shuffle the \textit{branch-to-grow} list \;
		\ForEach{\textit{branch-to-grow}}
		{
			\For{\textit{child} = 1 \textbf{to} 2}
			{
				Set branch angle $\alpha = \alpha_0\left\lbrace 1.0  + 0.1 * \rand\left(-1, 1\right) \right\rbrace $ \;
				Compute $\hat{\mathbf d_0}$ with Eq. \eqref{eq:motherdirection} \;
				\For{$j = 1$ \textbf{to} $N$}
				{
					Get the gradient of distance $\mathbf d^{grad}$ at $\mathbf r_{j-1}$ with Eq. \eqref{eq:graddistance} \;
					Compute segment direction $\mathbf d_j$ with Eq. \eqref{eq:branchgrowing} \;
					Create a new node by $\mathbf r_{j} = \mathbf r_{j-1} + l_s * \mathbf d_j$ \;
					Project node at $\mathbf r_{j}$ to surface with Eq. \eqref{eq:nodeproject} \;
					Collision detection with Eq. \eqref{eq:collision-criterion} \;
					\If{\textit{collision}}
					{
						Break \;
					}
				}
				\If{No \textit{collision}}
				{
					Add the new fully grown bracnh the \textit{branch-to-grow} list \;
				}
			Set branch angle $\alpha = - \alpha$ \;
			}
		}
	}
	\caption{Efficient network generation on arbitrarily complex surface.}
	\label{alg:network}
\end{algorithm}
%
\subsection{Nearest node search with CLL scheme}\label{sec:nodesearch} 
\begin{figure}[htb!]
	\centering
	\includegraphics[trim = 3cm 4cm 2cm 2.5cm, clip, width=.95\textwidth]{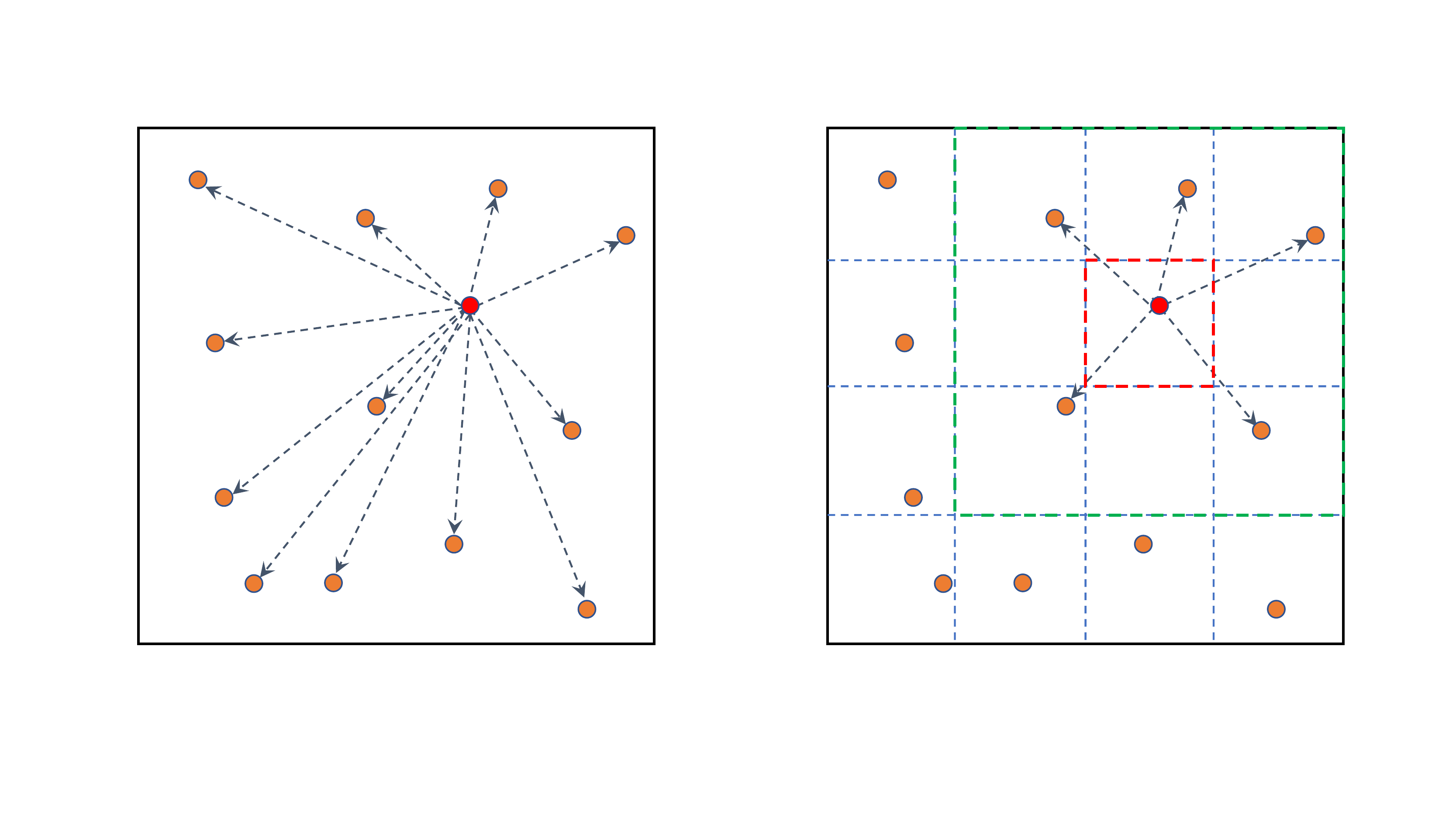}
	\caption{Schematic diagram for the CLL scheme for nearest node search. 
		The neighbor search of a single node can be conducted by searching all other nodes (left panel)or by dividing the domain into cells 
		with an length of at least the cutoff radius and searching the neighbors between the node 
		and all nodes in the same (red) and in the adjacent (green) cells. 
		Here, we consider the cutoff radius equals to $2.6 l_{seg}$ for consistency with the SPH framework. 
		(For interpretation of the references to color in this figure legend, the reader is referred to the web version of this article.)
	}
	\label{figs:cll}
\end{figure}

Different with the work of Costabal et al. \cite{costabal2016generating} 
where a $k-d$ tree binary structure is applied for nearest node search, 
we apply the CLL scheme which is widely used in the particle-based method for neighboring particles search. 
The CLL works by subdividing the whole computational domain into cells 
with an cell length greater than or equal to a specific cut-off radius as shown in Figure \ref{figs:cll}. 
The nodes are sorted into these cells and 
the neighbor search are conducted between nodes in the same or neighboring cells. 
The network generation process can be efficiently embedded into 
the CLL scheme by added newly created node into its corresponding cell 
without changing the whole data structure. 
Note that we set the cutoff radius equals to $2.6 l_{seg}$ similar to the one used 
for the $5th$-order Wendland smoothing kernel \cite{wendland1995piecewise} 
which is applied for all the simulations presented in this work. 

%
%
%
\section{Reduced-order SPH method}\label{sec:reducedoder}
In this section, 
we present a reduced-order SPH method for discretizing the one-dimensional monodomain equation 
in a linear structure in three-dimensional space, 
and then the corresponding neighboring particles search is fully described as follows.
\subsection{Reduced-order SPH method}\label{sec:reduced-order}
To model the electrical activation through the Purkinje network,  
we derive a reduced-order SPH method to discretize the one-dimensional monodomain equation. 
Considering that the electrical activation through the Purkinje network is characterized by one-dimensional wave propagation 
through three-dimensional space, 
we can drive a reduced-order SPH method by constraining the degree of freedom along one space dimension.  

In the full-order SPH method \cite{monaghan1992smoothed, lucy1977numerical, hu2006multi}, 
the kernel approximation of a continuous function $f\left( \mathbf r \right)$ reads
\begin{equation}\label{eq:kenelintegral}
	f \left( \mathbf r\right)   \approx \int_ \Omega  f \left( \mathbf r' \right)  W \left( \mathbf r-\mathbf r', h\right) d\mathbf r', 
\end{equation}
where $\Omega$ denotes the volume of the integral domain and 
$W(\mathbf{r}-\mathbf{r'}, h)$ the smoothing kernel function with smoothing length $h$ defining the support domain. 
By carrying out the integration of Eq. \eqref{eq:kenelintegral} along the to-be-reduced dimension, 
we can rewrite the approximation of $f\left( \mathbf r \right) $ as
\begin{equation}\label{eq:reducedkenelintegral}
	f \left( \mathbf r\right)   \approx \int_{\widehat{\Omega}}  f \left( \mathbf r' \right)  \widehat{W} \left( \mathbf r-\mathbf r', h\right) d\mathbf r', 
\end{equation}
where $\widehat{\Omega}$ denotes the to-be-reduced dimension space and 
$\widehat{W}$ the reduced-order kernel function. 
For solving one-dimensional problem in three-dimensional space, 
the integration space is reduced from three to one dimension. 
Therefore, 
a reduced-order fifth-order Wendland kernel \cite{wendland1995piecewise} reads
\begin{equation}
	\widehat W (q, h) =  \alpha
	\begin{cases}
		\left(1 + 2q \right) \left( 1 - 0.5q \right)^4  & \text{if} \quad 0\le q\le2 \\ 
		0 & \text{otherwise} 
	\end{cases},
\end{equation}
where $q = \left| \mathbf r-\mathbf r'\right|/h$ and the constant $\alpha$ is equal to $0$, $0$ and $3/4h$ in 
one, two or three dimensions, respectively. 
Note that the reduced-order kernel function has identical form with the full-order counterpart except 
different dimensional normalizing constant parameter, 
allowing the condition of unit can be satisfied in the reduced space. 

Following Ref. \cite{monaghan1992smoothed, zhang2021integrative}, 
the gradient of function $f$ can be approximated by 
\begin{equation}\label{grad-pe}
	\nabla f\left(\mathbf r \right)  \approx \int_{\widehat{\Omega}} \nabla f\left( \mathbf r' \right) \widehat{W} \left( \mathbf{r}-\mathbf{r'}, h\right)  d \left( \mathbf{r'}\right).
\end{equation}
Subsequently, 
it's not difficulty to derive the particle approximation of $\nabla f$ in reduced-order SPH strong form as
\begin{equation}\label{eq:reducedsphstronggrad}
	\nabla f_i = f_i \nabla 1 + \nabla f_i \approx   \sum_j V_j  \left(f_i - f_j \right)  \nabla_i \widehat{W}_{ij}, 
\end{equation}
and weak form 
\begin{equation}\label{eq:reducedsphweakgrad}
	\nabla f_i = \nabla f_i - f_i \nabla 1 \approx   - \sum_j  V_j  \left( f_i + f_j \right)  \nabla_i \widehat{W}_{ij} . 
\end{equation}
Here, 
$V$ is the particle volume, 
$f_i \equiv f(\mathbf{r}_i)$, 
$f_j  \equiv f(\mathbf{r}_j)$ and 
$\nabla_i \widehat W_{ij} \equiv \nabla _i \widehat W(r_{ij}, h)$ where $r_{ij} = | \mathbf{r}_{i} - \mathbf{r}_j|$. 
\subsection{Reduced-order SPH discretization for monodomain equation}\label{sec:reducedsph}
Following our previous work \cite{zhang2021integrative}, 
we employ the operator splitting method to decouple the monodomain equation into a PDE of 
\begin{equation}\label{eq:diffusion}
	S_d \quad : \quad  C_m \frac{\text{d}V_m}{\text{d}t}  = \nabla \cdot (\mathbb{D} \nabla V_m), 
\end{equation}
and two ODEs 
\begin{equation}\label{eq:ode-system}
	S_r \quad : \quad 
	\begin{cases}
		C_m \frac{\text{d}V_m}{\text{d}t}  = I_{ion}(V_m, w) \\
		\frac{\text{d}w}{\text{d}t} = g(V_m, w) 
	\end{cases},  
\end{equation} 
where $I_{ion}(V_m, w)$ and $ g(V_m, w)$ are defined by the Aliev-Panfilow model \cite{aliev1996simple}. 
Here, 
the operators $S_d$ corresponds to the diffusion step and $S_r$ the reaction step, 
and more details are referred to Ref. \cite{zhang2021integrative}. 
Then, 
we employ the 2nd-order Strang splitting \cite{strang1968construction} 
to approximate the solution of the monodomain equation 
from time $t$ to $t + \Delta t$ as
\begin{equation}\label{eq:spliting-2rd}
	V_m\left( t + \Delta t\right)  = S_r(\frac{\Delta t}{2}) \circ S_d(\Delta t) \circ S_r(\frac{\Delta t}{2}) V_m\left( t \right) ,
\end{equation}
where the $ \circ $ symbol separates each operator and indicates that an operator is applied to the following arguments. 
Similar to Ref. \cite{zhang2021integrative}, 
a reaction-by-reaction splitting operator with the quasi-steady-state (QSS) solver is applied for the operator $S_r$. 
As for the operator $S_d$, 
we apply the anisotropic SPH discretization within the reduced-order framework. 
Subsequently,  
the diffusion step can be solved by 
\begin{equation}\label{grad-laplace-v}
	S_d \quad : \quad \frac{\text{d}V_{m, i}}{\text{d}t} = \frac{2 d^P_{iso}}{C_m} \sum_{j} V^0_j \left(V_{m,i} - V_{m,j} \right) 
	 \frac{1}{|\mathbf{r}^0_{ij}|} \frac{\partial \widehat{W} \left( |\mathbf{r}^0_{ij}|, h \right)}  {\partial \left( |\mathbf{r}^0_{ij}|\right) } ,
\end{equation}
where $V_{m, i}$ and $V_{m, j}$denote the transmembrane potential of particle $i$ and $j$, respectively. 
\subsection{Network-based neighboring particle search}\label{sec:reducedsearch}
Having the network generation, 
the corresponding particle discretization can be initialized by converting each node to one particle and assigning the particle volume as $l_g^3$. 
To implement the particle interaction configuration, 
which consists of determining particle-neighbor lists and computing corresponding kernel weights and gradients, 
we introduce a network-based neighboring particle search scheme. 

\begin{figure}[htb!]
	\centering
	\includegraphics[trim = 1cm 5cm 1cm 4cm, clip, width=.95\textwidth]{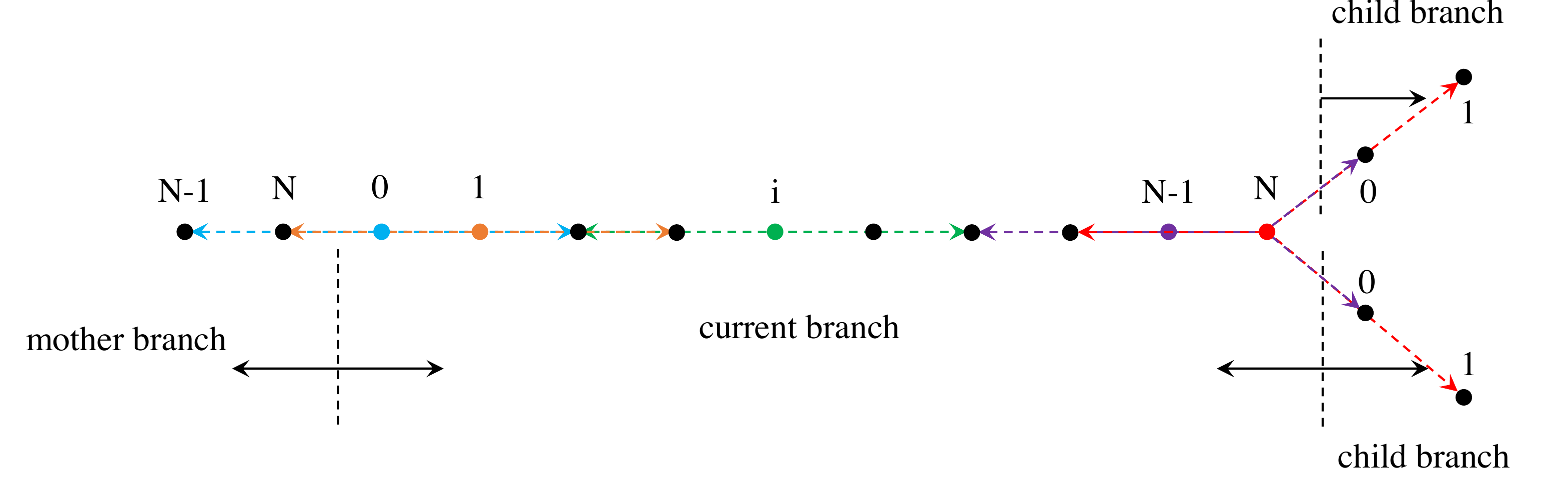}
	\caption{Schematic diagram for the network-based neighboring particle search scheme.
		The searching domain for typical particles are represented by the arrow line with the same color.
		(For interpretation of the references to color in this figure legend, the reader is referred to the web version of this article.)
	}
	\label{figs:nodeneighbor}
\end{figure}

The working principle of the network-based neighboring particle search scheme is that 
each particle can only has interactions with 
particles from its \textit{mother}, \textit{child} and current branches. 
In this case, 
the electrical activation resolved by the reduce-order SPH method is characterized by a propagation traveling from 
a \textit{mother} branch to its \textit{child} branches.  
Then, 
the network-based neighboring particles search scheme can be described as follows. 
As $N$ particles are converted from each branch, 
they can be categorized into five groups: 
(1) the particle $0$ has two neighboring particles from \textit{mother} branch with indexes $N$ and $N-1$ 
and another two from current branch with indexes $1$ and $2$;
(2) the particle $1$ has one neighboring particles from \textit{mother} branch with index $N$ 
and another three from current branch with indexes $0$, $2$ and $3$;
(3) the particle $N-1$ has one neighboring particle from each \textit{child} branch with index $0$ 
and another three form current branch with index $N-3$, $N-2$ and $N$;
(4) the particle $N$ has two neighboring particles from each \textit{child} branch with indexes $0$ 
and $1$, and another two from current branch with indexes $N-2$ and $N-1$;
(5) other particle $i \in [2, N-2]$ has four neighboring particles from current branch with indexes $i-2$, $i-1$, $i+1$ and $i+2$. 
Here, 
the index denotes the local index of current branch. 
The present neighboring particle search scheme is illustrated in Figure \ref{figs:nodeneighbor}. 
Following the present process, 
each particle has $4\sim5$ neighboring particles compatible 
with the smoothing length $h = 1.3dp$ with $dp$ denoting the particle spacing, 
widely applied for Wendland kernel function. 
%
%
%
\section{Multi-order SPH method}\label{sec:multiorder}
In this section, 
the multi-order SPH method is presented to solve the MM coupling problem 
between the Purkinje network and the myocardium.  
To optimize the computational efficiency, 
a multi-time stepping scheme is introduced for 
the time integration of multi-physics applications. 
\subsection{Multi-order coupling paradigm}\label{sec:multiorder-coupling}
\begin{figure}[htb!]
	\centering
	\includegraphics[trim = 1cm 3cm 1cm 3cm, clip, width=.95\textwidth]{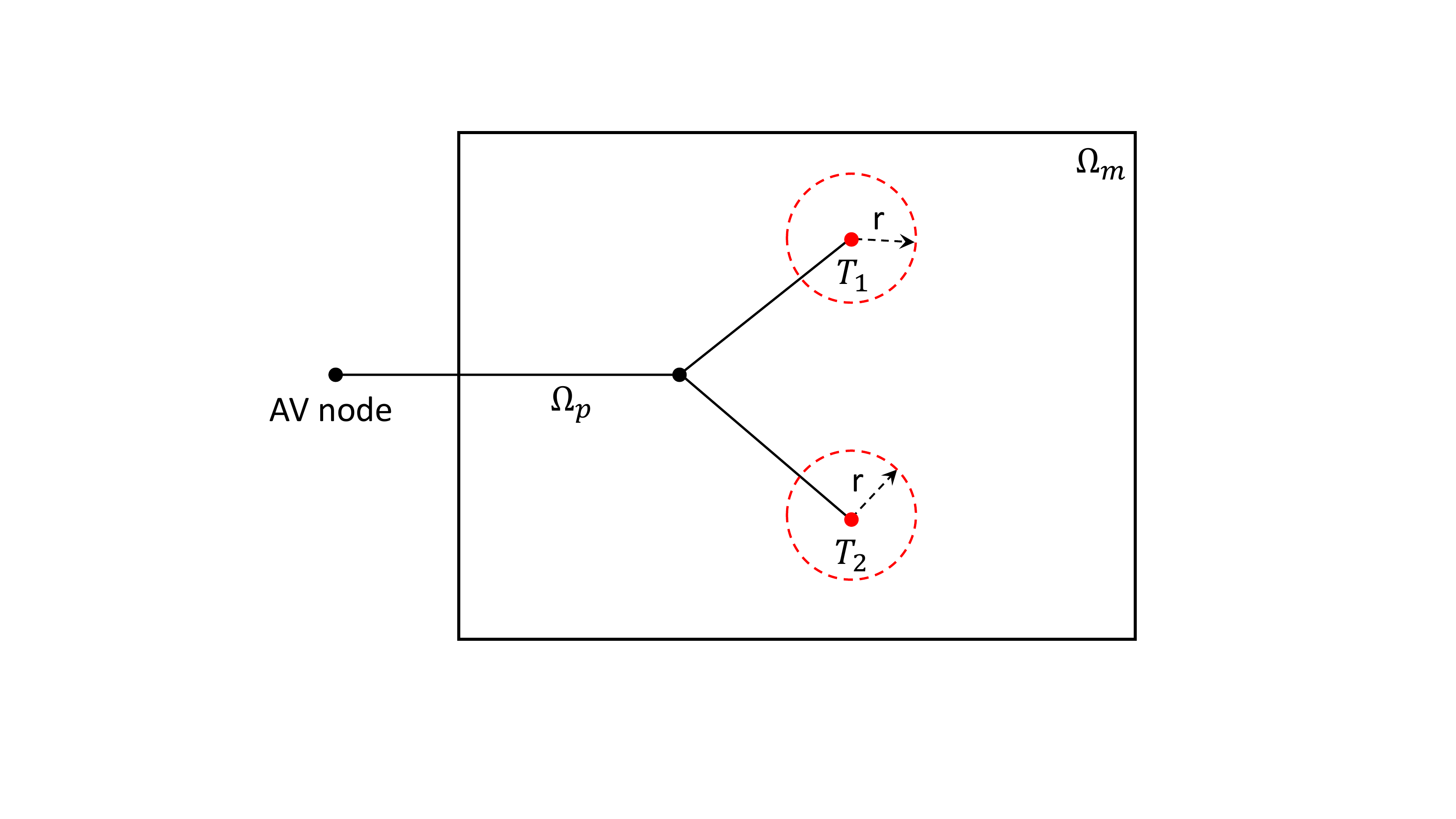}
	\caption{Schematic diagram of a simple myocardial domain $\Phi_m$ with a generic network $\Phi_p$. 
		Here, 
		the generic network consists of the AV node and two terminal nodes $T_1$ and $T_2$ denoting the PKJs which 
		have the influence of spherical region with a radius of $r$ centered in the node.
		(For interpretation of the references to color in this figure legend, the reader is referred to the web version of this article.)
	}
	\label{figs:pkjmyo}
\end{figure}

As a subendocardial network, 
the Purkinje system characterized by a high conduction velocity 
and is isolated from the muscle, 
except their PKJs which located on the endocardium. 
Through these PKJs, 
the electrical signal enters the myocardium and gives rise a coupling nature of potential propagation. 
In the present reduced-order SPH method, 
the PKJs are represented by terminal particles converted from terminal nodes from branches which are terminated due to collision. 
Then, 
the terminal particles act as current sources of the full-order myocardium particles located inside their influence region. 
Similarly to Ref. \cite{vergara2016coupled}, 
the influence region is modeled by a spherical region of radius $r$ centered at the PKJ as shown in Figure \ref{figs:pkjmyo}. 
The influence region can be recovered straightforwardly by the SPH method due to the fact that 
each particle has a spherical support domain determined by the cutoff radius. 

For the coupling between reduced-order and full-order structures, 
namely the Purkinje network and the myocaridum, 
the particles of each coupling pairs are in different level of reduced order. 
The basic idea is first to transfer the lower-order particles into the high-oder particles. 
Therefore, the coupling is transferred into single-order and 
the interpolation and interface particle fluxes can be computed. 
One important property for such coupling in the present project is that 
the material properties of the coupling structures are not very different. 
Therefore, 
coupling between the Purkinje network and the myocardium 
where terminal particles take the roles of excitation sources can be conducted straightforwardly. 

Consider the MM coupling model, 
the monodomain equation for the myocardium particles located in the influence region of 
the terminal particles of the Purkinje network
can be rewritten as 
\begin{equation}\label{eq:mmcoupling}
	C_{m, a} \frac{\text d V_{m, a}}{\text d t} = \nabla^0 \cdot  \left(\mathbb D \nabla^0 V_{m, a}\right)  + I_{ion} + I_a^{M:P}, 
\end{equation}
where superscript $a$ represents the myocardium particles and 
$I_a^{M:P}$ denotes the current flux from network to myocardium. 
For the myocardium particles located in the influence region of terminal particles, 
the current flux can be obtained by 
\begin{equation}\label{eq:mmflux}
	 I_a^{M:P} = 2 \left(d^M_{iso} + d^P_{iso}\right)  \sum_i V^0_i \left(V_{m,a} - V_{m,i} \right)  \frac{1}{|\mathbf{r}^0_{ai}|} \frac{\partial W \left( |\mathbf{r}^0_{ai}|, h \right)}  {\partial \left( |\mathbf{r}^0_{ai}|\right) } ,
\end{equation}
where subscript $i$ represents the terminal particles. 
Here, 
we assume that the network and the myocardium particles have identical smoothing length, 
implying that $l_{seg} = dp_0$ with $dp_0$ the initial myocardium particle spacing. 
Then, 
the neighboring terminal particles from the Purkinje network 
of a myocardium particle can be searched though the CLL scheme summarized in the previous Section \ref{sec:nodesearch}. 
Subsequently, 
the influence region of PKJ represented by the corresponding terminal particle is resolved by a sphere with a radius of $2.0 h$. 
%
\subsection{Multi-time stepping scheme}\label{sec:coupling} 
For multi-physics problems, 
the time integration scheme plays a key role in determining the computational performance. 
One widely applied approach is the single-time stepping scheme 
which chooses the minimal time step size of all sub-systems as the applied one
and this scheme may induces excessively computational efforts as demonstrated in Ref. \cite{zhang2021multi}. 
On the other hand, 
a multi-time stepping scheme can be introduced to optimize the computational efficiency. 
For the numerical study of cardiac electromechanics coupling problem of the myocardium with inclusion of the Purkinje network, 
we have three criteria to determine the time step size for stable time integration. 
More precisely, 
the time step size $\Delta t^P_d $ for solving the monodomain equation in the Purkinje network is given by 
\begin{equation}\label{eq:dt_p}
	\Delta t^P_d = \frac{1}{2d} \left( \frac{h^2}{d^P_{iso}} \right),
\end{equation}
and the $\Delta t^M_d $ for solving the monodomain equation in the myocardium is determined by 
\begin{equation}\label{eq:dt_m_d}
	\Delta t^M_d = \frac{1}{2d}\left( \frac{h^2}{|\mathbb{D}|} \right). 
\end{equation}
Here, 
$d$ is the dimensionality and $|\mathbb{D}|$ the trace of the diffusion tensor. 
As for the mechanical response of the myocardium, 
the time step size $\Delta t^M_{m}$ is defined as 
\begin{equation}\label{eq:dt_m_m}
	\Delta t^M_{m}   =  0.6 \min\left(\frac{h}{c + |\mathbf{v}|_{max}},
	\sqrt{\frac{h}{|\frac{\text{d}\mathbf{v}}{\text{d}t}|_{max}}} \right), 
\end{equation}
where $c$ denotes the speed of sound for passive mechanical response. 
Also note that we assume the three sub-systems has identical smoothing length. 

Subsequently, 
we have three time step sizes, 
namely $\Delta t^P_d$, $\Delta t^M_{d}$ and $\Delta t^M_{m}$. 
In general, 
the time step size $\Delta t^M_{m}$ for mechanical response of the myocardium has the minimal value 
as it is dominated by the speed of sound. 
Also, 
$\Delta t^P_d$ is smaller than $\Delta t^M_{d}$ as the electrical activation travels more rapidly through the Purkinje network 
than that in the myocardium.
With that $\Delta t^M_d > \left\lbrace \Delta t^P_{d}, \Delta t^M_{m}\right\rbrace $, 
we introduce the following multi-time stepping method. 
Other than choosing the minimal step size as the single time step for all the subsystems, 
we carry out the integration of monodomain equation in the myocardium with $\Delta t^M_d $, 
during which 
$\varkappa = [\frac{\Delta t^M_d}{\Delta t^F_d}] + 1$ times integration of monodomain equation in the Purkinje network 
and
$\kappa = [\frac{\Delta t^M_d}{\Delta t^M_m}] + 1$ times integration of active mechanical response in the myocardium 
are conducted simultaneously. 
Note that $[\cdot ]$ represents the integer operation. 
The detailed algorithm for the present multi-time stepping algorithm is presented in Algorithm \ref{alg:multistepping}. 
%
\begin{algorithm}[htb!]
	Setup the total simulation time $T$ \;
	\While{$t < T$}
	{
		Compute $\Delta t^M_d$ with Eq. \eqref{eq:dt_m_d} \;
		Integrate the monodomain qutaion in myocardium with time step size $\Delta t^M_d$ \;
		Set $t^P_{sum} = 0$ \;
		\While{$t^P_{sum} < \Delta t^M_d$}
		{
			Comput $\Delta t^P_d$ with Eq. \eqref{eq:dt_p} \;
			Integrate the monodomain qutaion in the Purkinje network with time step size $\Delta t^P_d$ \;
			Update sub-integration time with $t^P_{sum} = t^P_{sum} + \Delta t^P_d$ \;
		}
		Set $t^M_{sum} = 0$ \;
		\While{$t^M_{sum} < \Delta t^M_d$}
		{
			Compute $\Delta t^M_m$ with Eq. \eqref{eq:dt_m_m} \;
			Integrate the mechanical equation in myocardium with time step size $\Delta t^P_m$ \;
			Update sub-integration time with $t^M_{sum} =t^M_{sum} + \Delta t^M_m$ \;
		}
		Update the total integration time with $t  = t + \Delta t^M_d$ \;
	}
	Finalize the Computation \;
	\label{alg:multistepping}
	\caption{The detailed procedure for the time integration using the proposed multi-time stepping scheme
		for solving electromechanics problem of the myocardium with inclusion of the Purkinje network.}
\end{algorithm}
%
%
%
%
\section{Numerical examples}\label{sec:examples}
In this section, 
we present a set of numerical results with the aim of assessing the computational efficiency, 
accuracy and versatility of the proposed methods for numerical study of electromechanics coupling problem of the myocardium 
with inclusion of the Purkinje network. 
First of all, 
we assess the computational efficiency of the proposed algorithm for network generation in comparison with the one presented in Ref. \cite{costabal2016generating}. 
After this preliminary step, 
we validate the reduced-order SPH method for resolving the electrical activation through the network. 
Then, 
we consider an academic test with simplified myocardium coupled with a generic network to demonstrate the accuracy of 
the proposed multi-order SPH method with MM coupling model. 
Having the validations, 
we apply the present method to approximate the electrophysiology and electromechanics problems of the realistic left ventricle 
with inclusion of the Purkinje network. 

In all the following examples,
the full-order $5th$-order Wendland smoothing kernel function with the smoothing lengths $h = 1.3dp$ 
with $dp$ denoting the initial particle spacing  is applied for the resolving 
the electrical activation and active mechanical response in the myocardium, 
and more details of the numerical algorithms are referred to our previous work of Ref. \cite{zhang2021integrative}.
\subsection{Computational efficiency of network generation}\label{sec:efficiency} 
To assess the computational performance for network generation, 
we analyze the total CPU time of the present algorithm and the one developed by Costabal et al. \cite{costabal2016generating} 
where a k-d tree scheme is applied for nearest node search and the node projection is conducted by directly searching the neighboring triangles. 
In this work, 
the computations are carried out on an Intel Xeon CPU E5-2620 v3 2.40GHz desktop computer with 64GB RAM and Scientific Linux system (7.9). 
Note that the present algorithm is implemented in our open-source SPHinXsys library which is available at \url{https://www.sphinxsys.org} \cite{zhang2021sphinxsys} 
and the algorithm of Costabal et al. \cite{costabal2016generating} 
is available in the repository at \url{https://github.com/fsahli/fractal-trees}. 
\begin{figure}[htb!]
	\centering
	\includegraphics[trim = 1mm 1mm 1mm 1mm, clip, width=.95\textwidth]{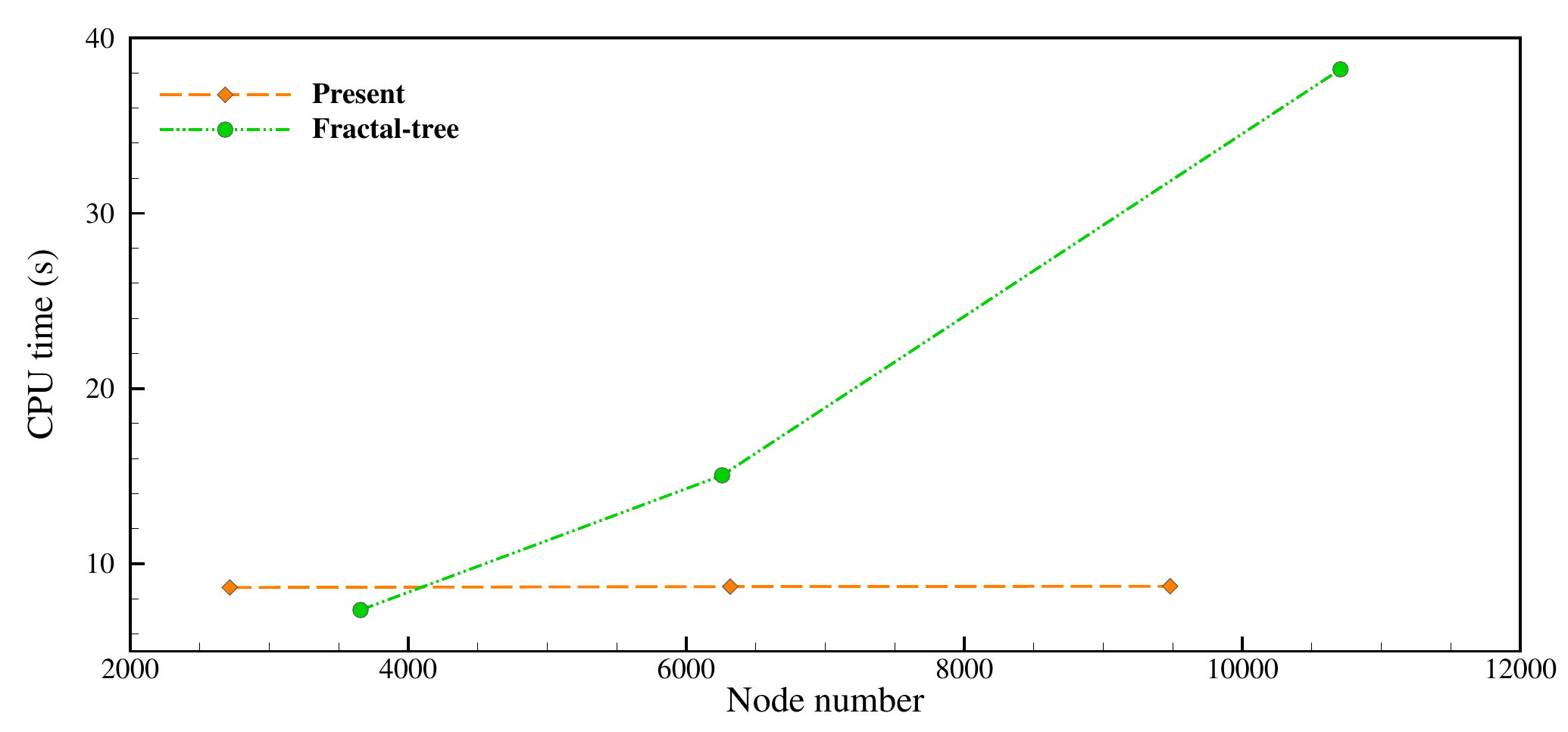}
	\caption{The total CPU time of network generation for different node number by using the present algorithm and 
		the one developed by Costabal et al. \cite{costabal2016generating}. 
		(For interpretation of the references to color in this figure legend, the reader is referred to the web version of this article.)
	}
	\label{figs:cpu-time}
\end{figure}

Figure \ref{figs:cpu-time} shows the CPU time with respect to network generation with different number of node by 
using the present algorithm and the one proposed by Costabal et al. \cite{costabal2016generating}.
It can be noted that the computational cost of the present algorithm exhibits invisible increase as the increase of the node number. 
However, 
the algorithm of Ref. \cite{costabal2016generating} shows linear increase of the computational cost as the increase of the node number due to 
the fact that neighboring triangle search is essential for every newly created node. 

To rigorously analyze the computational performance of the present algorithm, 
we decompose the total CPU time into two parts, 
i.e., the pre-processing and the network generation. 
The pre-processing part includes the parsing of polygon mesh and the construction of the corresponding level-set field,  
and the network generation part mainly consists of the iteration process for branch growth. 
Table \ref{tab:cpu-analysis} reports the computational cost 
for different parts of the present algorithm for network generation with different node number. 
The pre-processing part accounts for up to $99\%$ of the total computational time and 
the network generation part takes less than $1\%$. 
This analysis explains why the present algorithm can effectively improve the computational efficiency by 
avoiding the neighboring triangle search for every newly created node. 
\begin{table}[htb!]
	\centering
	\caption{The analysis of computational cost for different parts of the present algorithm for network generation. 
		The total computational time is decomposed into pre-processing, 
		which consists of the paring of a polygon mesh and the construction of the level-set field, 
		and the network generation part corresponds to iteration process of branch growth.
	}
	\begin{tabular}{cccc}
		\hline
		Number of nodes   & \begin{tabular}{c} Total  \\ CPU time(s) \end{tabular}  &  \begin{tabular}{c} Pre-processing  \\ CPU time(s) \end{tabular}&  \begin{tabular}{c} Network generation  \\ CPU time(s) \end{tabular} \\ 
		\hline
		$2716$			& $8.6423$& $8.6277$     &$0.0146$  \\
		
		$6316$ 			& $8.6929$& $8.6588$     & $0.0341$ \\
		
		$9480$	 		& $8.7113$& $8.6593$     & $0.052$ \\
		
		\hline	
	\end{tabular}
	\label{tab:cpu-analysis}
\end{table}
%
\subsection{Transmembrane potential propagates through a myocardium fiber}\label{sec:fibre} 
To quantitively address the accuracy of the proposed reduced-order SPH method 
for resolving the one-dimensional electrical activation in three-dimensional structure, 
a well-established numerical test 
where the transmembrane potential propagates through a myocardium fiber is studied in the part. 
The fiber is considered as linear tissue with the length of $L = 20 \text{mm}$, 
and the tissue is assumed to have uniform capacity of $C_m = 1.0$ 
and isotropic conductivity $d = 0.1 \text{mm}^2 / \text{ms}$. 
Following Refs. \cite{patelli2017isogeometric, nitti2021curvilinear}, 
the ionic current is modeled by the Aliev-Panfilow model \cite{aliev1996simple} 
with the constant parameters given in Table \ref{tab:ap-1}. 
To activate the depolarization, 
a stimulus is applied at the discrete level by forcing the transmembrane potential $V_m$ associated to the outermost left particles to 
$V_m = 1.0$ for time interval $t \in [0, 0.5] \text{ms}$, 
allowing the excitation of a traveling action potential wave rightwards. 
Note that similar test with considering planar wave propagation over a rectangular slab 
was conducted in Refs. \cite{patelli2017isogeometric} where the numerical result is available for quantitative comparison. 
To discretize the fiber, 
total number of $50$ particles are applied in the present simulation. 
\begin{table}[htb!]
	\centering
	\caption{Transmembrane potential propagates through a fiber: Parameters for the Aliev-Panfilow model \cite{aliev1996simple}. }
	\begin{tabular}{cccccc}
		\hline
		k	& a & b  & $\epsilon_0$ & $\mu_1$ & $\mu_2$  \\
		\hline
		8.0	& 0.15   & 0.15      & 0.002 & 0.2 & 0.3  \\
		\hline	
	\end{tabular}
	\label{tab:ap-1}
\end{table}

\begin{figure}[htb!]
	\centering
	\includegraphics[trim = 2mm 2mm 2mm 2mm, clip, width=\textwidth]{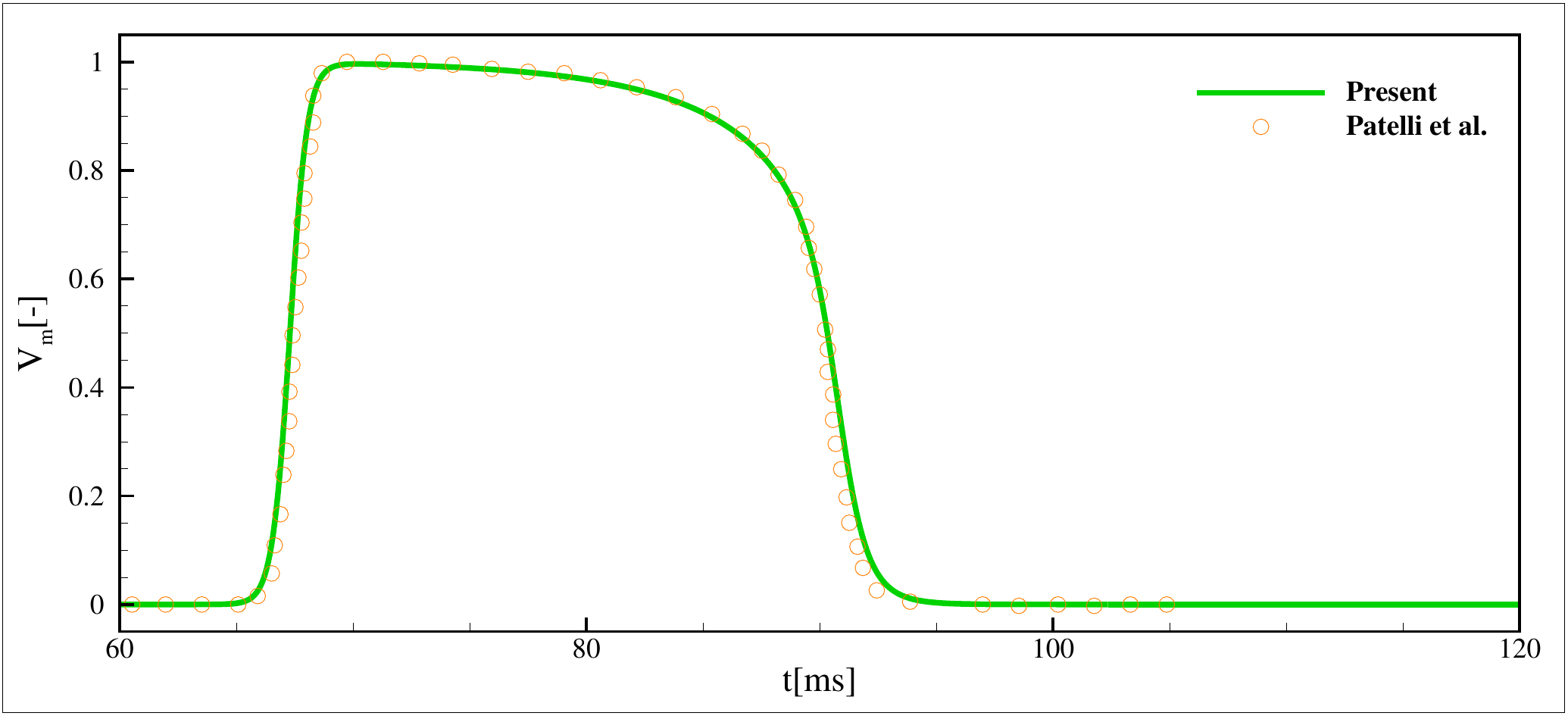}
	\caption{Transmembrane potential propagates through a fiber: 
		The time evolution of the transmembrane potential $V_m$ (green line) 
		superposed with the results reported by Patelli et al. \cite{patelli2017isogeometric} (orange dots).
		Note that the dimensional counterpart of the transmembrane potential $V_m$ 
		can be recovered by applying the transformation $V_m (\text{mV}) = -80 + 100V_m (-)$. 
		(For interpretation of the references to color in this figure legend, the reader is referred to the web version of this article.)
	}
	\label{figs:depolarization}
\end{figure}

Figure \ref{figs:depolarization} reports the predicted evolution profile of the transmembrane potential, 
and its comparison with that  reported by Patelli et al. \cite{patelli2017isogeometric}. 
It is observed that in accordance with the previous numerical estimation \cite{patelli2017isogeometric} 
and experimental observation \cite{franzone2014mathematical}, 
the quick propagation of the stimulus through the fiber 
and the slow decrease in the transmembrane potential after a plateau phase 
are well predicted by the present reduced-order SPH method. 
The level of agreement noted in the comparison with that of Patelli et al. \cite{patelli2017isogeometric} 
suggests that the present method can accurately resolve the one-dimensional electrical activation in a three-dimensional structure, 
and it provides result consistent with those from other state-of-the-art solvers \cite{patelli2017isogeometric, colli2012mathematical}. 
%
\subsection{Cubiod myocardium with inclusion of a generic network}\label{sec:couplingvalidation} 
\begin{figure}[htb!]
	\centering
	\includegraphics[trim = 1cm 10cm 4cm 5cm, clip, width= \textwidth]{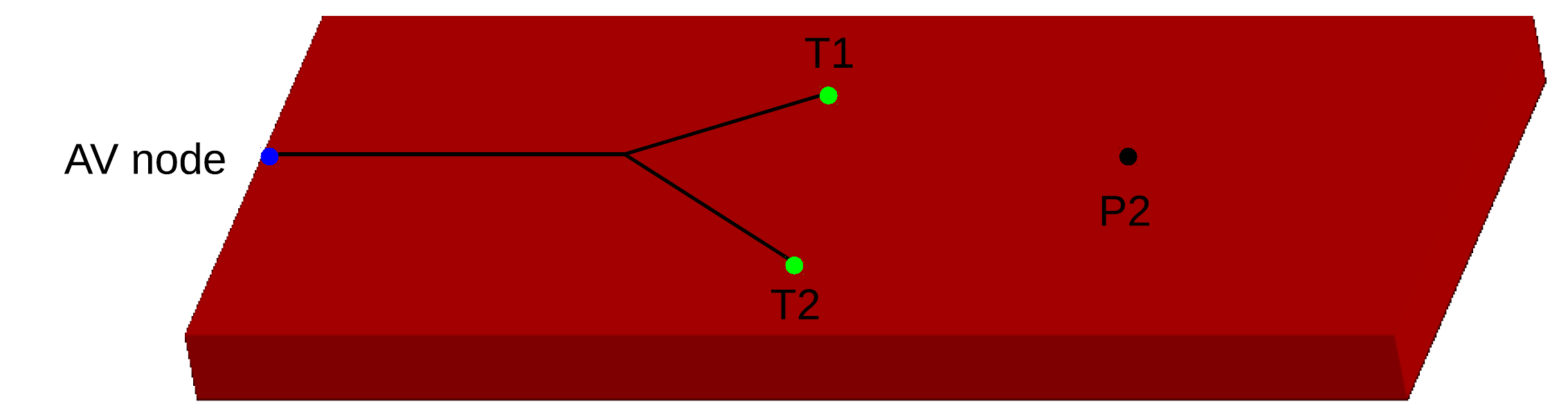}
	\caption{Cubiod myocardium with a generic Purkinje network: 
		The myocardial geometry is simplified as a slab with length $l_0 = 40 \text{mm}$, height $h_0 = 1 \text{mm}$ and width $w_0 = 20\text{mm}$, 
		and the network consists of three branches. 
		The transmembrane potential is initialized at the AV node, 
		travels through the network and then enters the myocardium through two PMJs, 
		i.e., $T1$ and $T2$. 
		The potential probe $P1$ located at the $(0.75 l_0, h_0, 0.5 w_0)$ where two wave front interacting with each other 
		is applied to probe the potential profile for validating the coupling paradigm. 
	}
	\label{figs:myo-network-setup}
\end{figure}

In this section, 
we validate the proposed multi-order SPH method. 
Following Ref. \cite{vergara2016coupled}, 
we consider potential propagation in simplified cubiod myocardium with a generic Purkinje network. 
The myocardium is simplified as a slab with length $l_0 = 40 \text{mm}$, height $h_0 = 1 \text{mm}$ and width $w_0 = 20\text{mm}$, 
and is modeled by orthogonal material with the fiber and sheet directions parallel to the global coordinates. 
The Purkinje network consists of three branches with two PKJs 
interacting with the myocardium as shown in Figure \ref{figs:myo-network-setup}. 
Note that the generic network lied on top of the cubiod domain, 
similar to physiological situation where the Purkinje fibers are located beneath the endocardium. 
For the transmembrane potential propagation, 
we apply a stimulus at the AV node, 
as shown in Figure \ref{figs:myo-network-setup}, 
allowing the potential travels through the network and enters the myocardium at two PKJs, i.e., $T1$ and $T2$. 
For validating the coupling paradigm. , 
the potential profile in the myocardium is probed at point $P1$ located at $(0.75 l_0, h_0, 0.5 w_0)$. 
The Aliev-Panfilow model \cite{aliev1996simple} is applied with the constant parameters given in Table \ref{tab:ap-1} 
for both the myocardium and the network.
 The diffusion coefficients for myocardium are set as  
 $d^M_{iso} = 0.1 \text{mm}^2 / \text{ms}$ and $d^M_{ani} = 0.01 \text{mm}^2 / \text{ms}$, 
 and for network $d^P_{iso} = 0.1\text{mm}^2 / \text{ms}$. 
 To discretize the system, 
 the initial particle spacing is set as $dp_0 = h_0 / 5$ for the myocardium and $l_{seq} = dp_0$ for the network. 

\begin{figure}[htb!]
	\centering
	\includegraphics[trim = 2mm 1cm 2mm 5mm, clip, width= \textwidth]{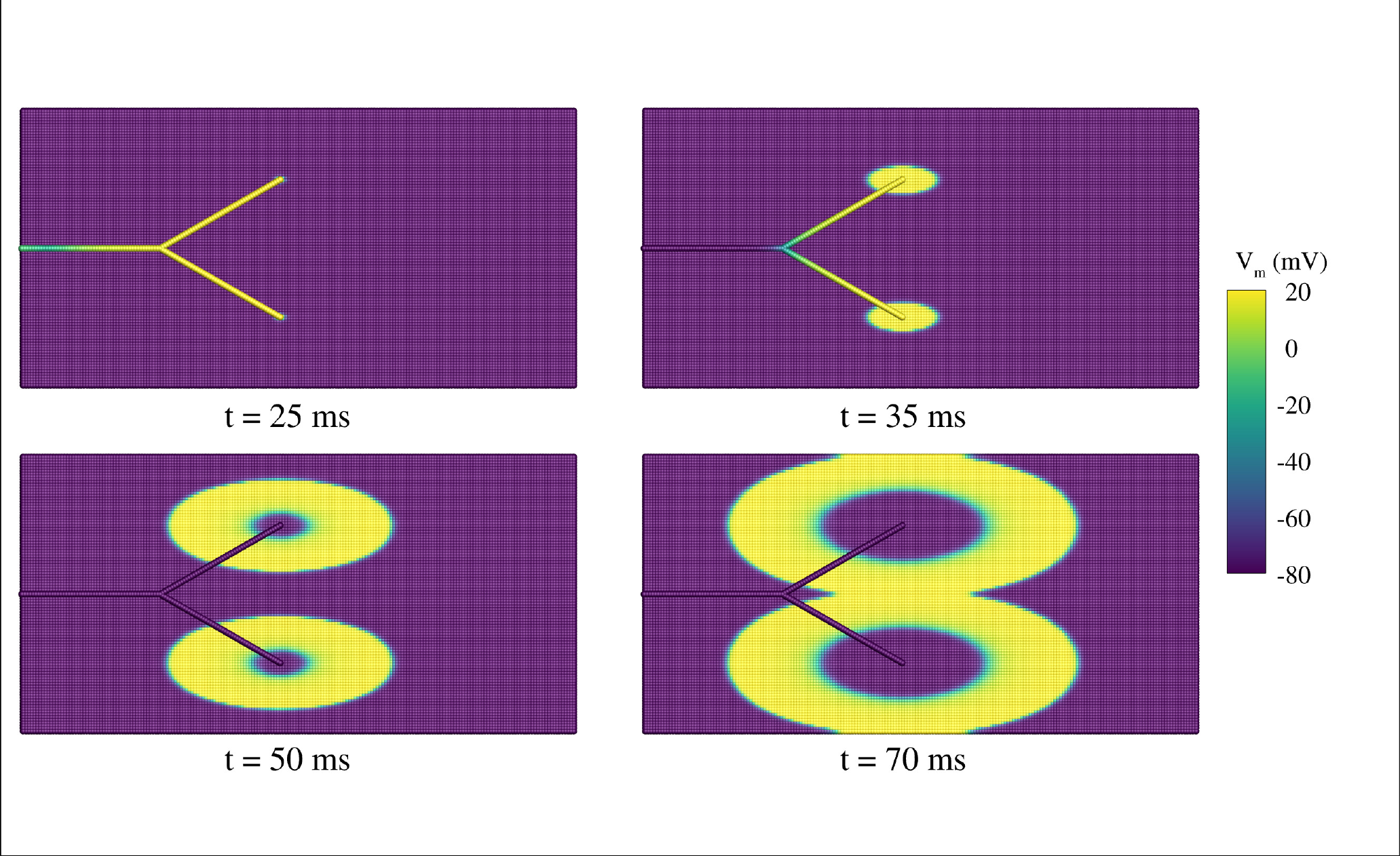}
	\caption{Cubiod myocardium with a generic Purkinje network: 
		Transmembrane potential at different temporal instants. 
		(For interpretation of the references to color in this figure legend, the reader is referred to the web version of this article.)
	}
	\label{figs:myo-network}
\end{figure}

Figure \ref{figs:myo-network} reports the transmembrane potential traveling through the network 
and in the myocardium at different time instants. 
As expected, 
the potential first travels through the network starting at the AV node, 
and then enters the myocardium through the PMJs, 
i.e., $T1$ and $T2$, 
allowing the activation of the two fronts in the myocardium. 
It is worth noting that the present multi-order coupling method demonstrates 
its ability to model in a proper way the collision of two fronts 
as observed in the frame at $t = 70 \text{ms}$. 

\begin{figure}[htb!]
	\centering
	\includegraphics[trim = 2mm 2mm 2mm 2mm, clip, width=\textwidth]{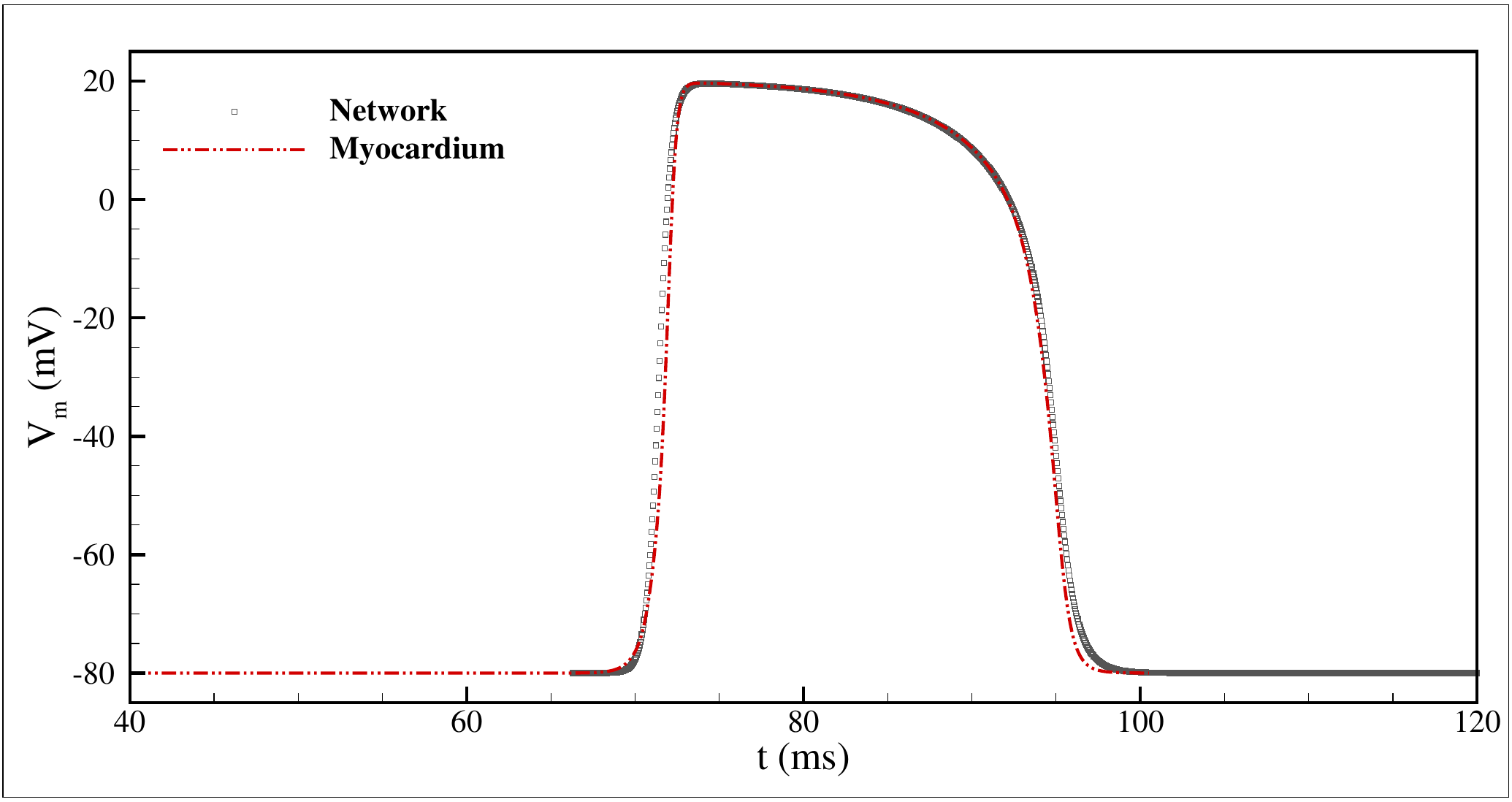}
	\caption{Cubiod myocardium with a generic Purkinje network: 
		The time evolution of the transmembrane potential $V_m$ (red line) observed at $P1$ 
		superposed with the potential profile of that through the network (black dots). 
		(For interpretation of the references to color in this figure legend, the reader is referred to the web version of this article.).
	}
	\label{figs:myo-network-data}
\end{figure}

To further investigate the accuracy of the present coupling method, 
Figure \ref{figs:myo-network-data} presents the time evolutions of the transmembrane potential 
at point $P1$ located at the myocardium 
and that through the network. 
The evolution of the simulated transmembrane potential suggests that the electrical signal activated through the PMJs 
traveling in the myocardium has the identical profile with that through the network. 
Not that an instantaneous activation of the myocardium is applied in the present coupling strategy 
and the delay in the normal propagation corresponding to the time necessary to excite the myocardial cells \cite{mendez1970propagation}
will be taken into consideration the the future work. 

%
\subsection{Left ventricle with inclusion of the Purkinje network}\label{sec:pkj-lv} 
\begin{figure}[htb!]
	\centering
	\includegraphics[trim = 1mm 1mm 1mm 5mm, clip, width= 0.85\textwidth]{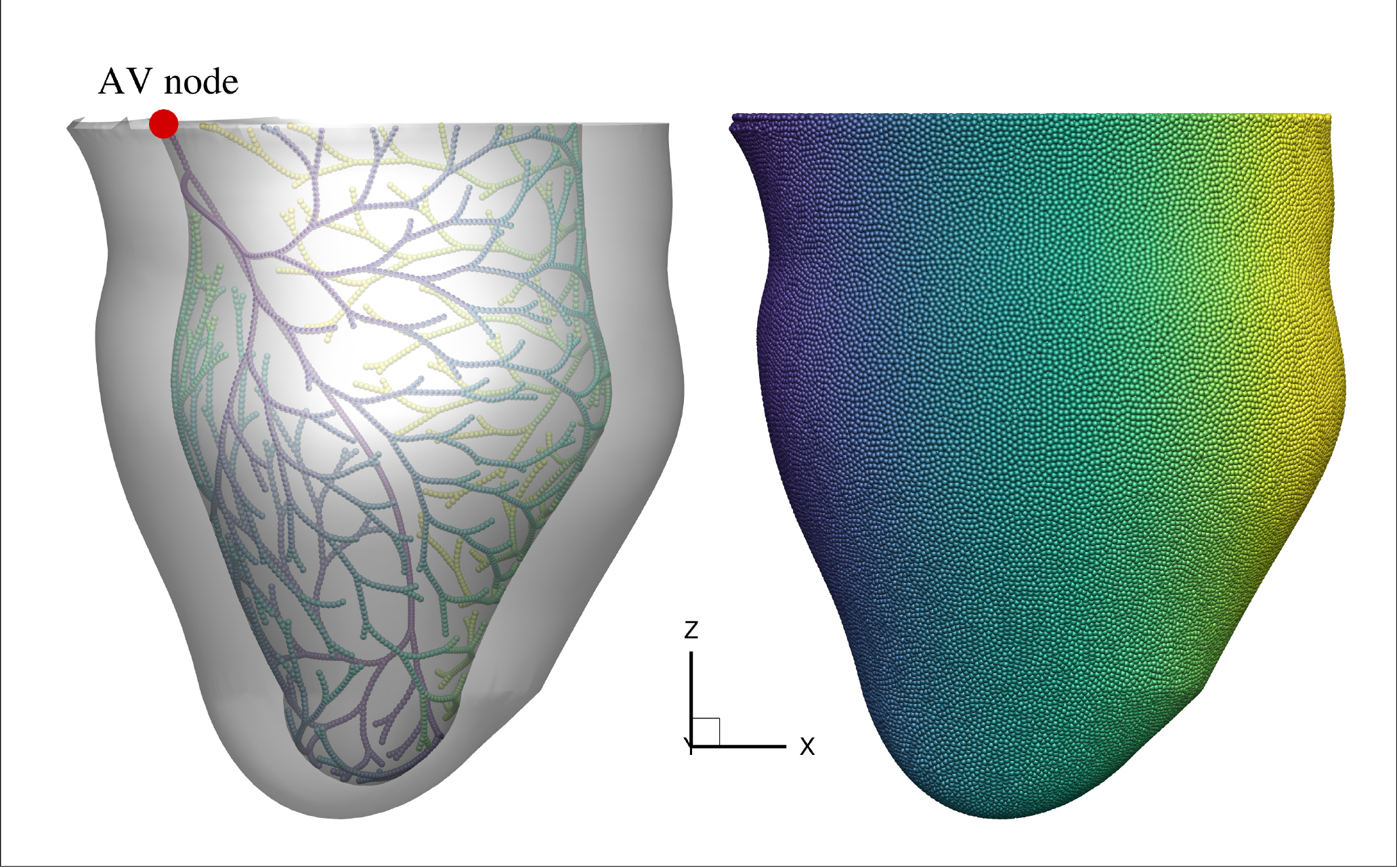}
	\includegraphics[trim = 1mm 1.5cm 1mm 5mm, clip, width= 0.85\textwidth]{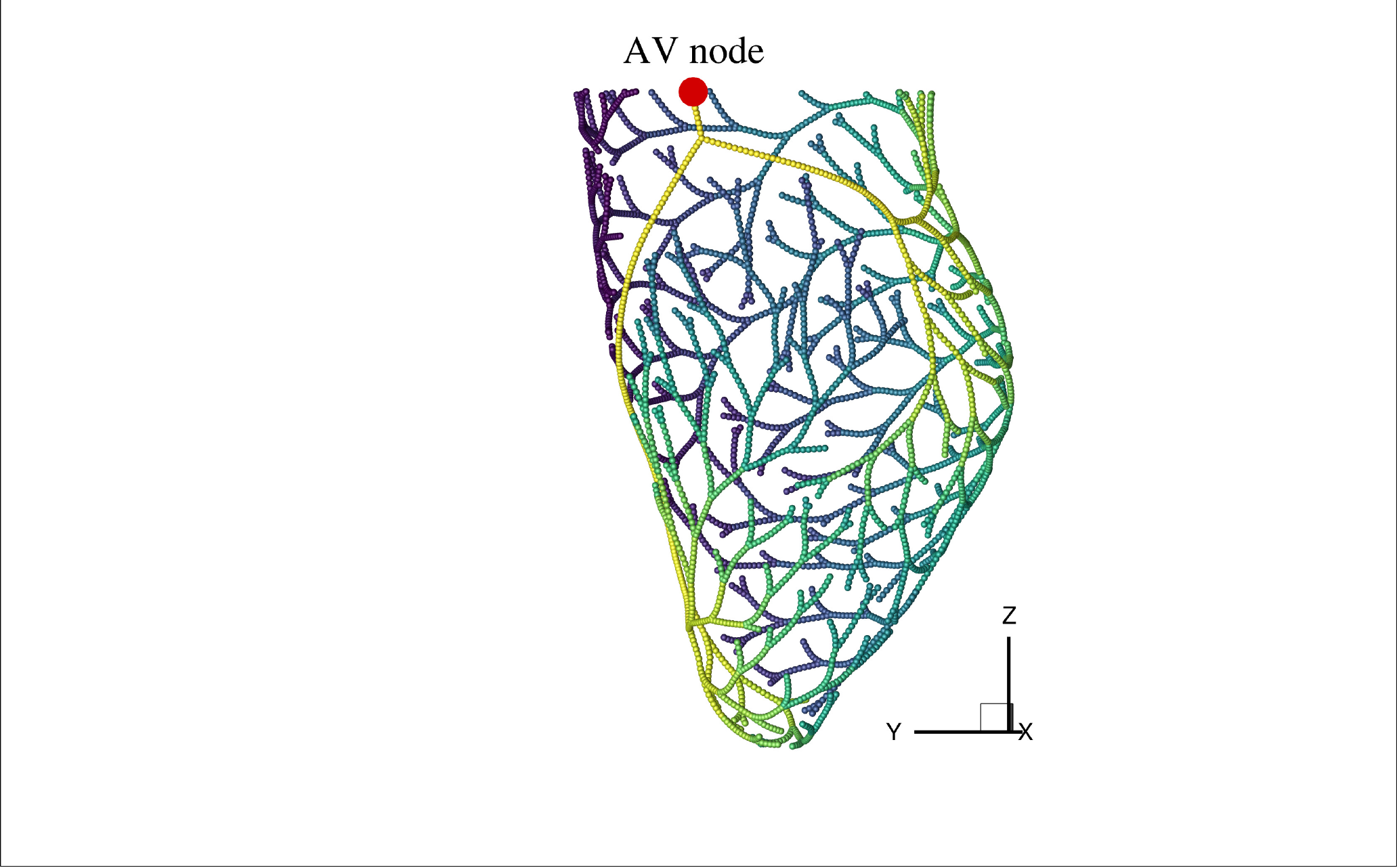}
	\caption{Left ventricle with the Purkinje network: 
		Representations of the left ventricle (upper left panel), 
		the particle model for the left ventricle (upper right panel) and the Purkinje network (bottom panel), respectively.
		Here, 
		the Purkinje network consists of $3888$ reduced-order particles and $259$ of which are terminal particles representing PKJs, 
		and the left ventricle is represented by $167209$ full-order particles.  
	}
	\label{figs:pkj-lv-setup}
\end{figure}

To demonstrate the versatility of the present multi-order SPH method in realistic cardiac modeling, 
we consider the transmembrane potential propagation and the corresponding excitation-contraction in the 
realistic left ventricle with inclusion of the Purkinje network. 
The left ventricle model applied herein was presented by Gao et al. \cite{gao2017coupled} 
where a cardiac magnetic resonance (CMR) study was performed on a healthy volunteer. 
As noted by Ref. \cite{gao2017coupled}, 
this study was approved by the local NHS Research Ethics Committee, and written informed consent was obtained before the CMR scan. 
Then, the left ventricle geometry and function were imaged with conventional short-axis and long-axis cine images whose parameters are referred to Ref. \cite{gao2017coupled}. 
From images at early-diastole, 
the left-ventricle geometry can be constructed using SolidWorks as shown in Figure \ref{figs:pkj-lv-setup}. 

Having the left ventricle geometry, 
we can generate a reduced-order particle model for the Purkinje network on the endocardial surface 
by using the present network generation algorithm.  
Figure \ref{figs:pkj-lv-setup} (bottom panel) illustrates the resulting Purkinje network, 
which consists of $518$ branches and $259$ of which are terminal ones. 
In the reduced-order SPH model, 
the network is represented by $3888$ reduced-order particles and $259$ of 
which are terminal particles which are interacting with the myocardium particles as PKJs. 
For generating the full-order particle model for the left ventricles, 
we apply the CAD-BPG mehtod proposed in our previous work \cite{zhu2021cad}. 
Figure \ref{figs:pkj-lv-setup} (left panel) shows the particle distribution for the left ventricle 
and it can be noted that an isotropic particle configuration is obtained and the geometry surface is reasonably well prescribed. 
Having the particle initialization of the left ventricle, 
the fiber and sheet reconstructions are conducted following Ref. \cite{zhang2021integrative, quarteroni2017integrated}. 

For both electrophysiology and electromechanics study, 
we consider three cases including, 
(i) the physiological healthy excitation where the unique source for the Purkinje network is the AV node 
and the unique sources for the myocardium were the PKJs;
(ii) the pathological excitation, namely the Wolff-Parkinson-White syndrome, 
which is characterized by an extra muscular intramyocardial source in addition to the AV node; 
(iii) the free-pulse excitation without the Purkinje network where the unique muscular source for the myocardium were located close to the AV node. 
The Aliev-Panfilow model \cite{aliev1996simple} is applied with the constant parameters given in Table \ref{tab:ap-2}. 
\begin{table}[htb!]
	\centering
	\caption{Left ventricle with the Purkinje network: Parameters for the Aliev-Panfilow model \cite{aliev1996simple}. }
	\begin{tabular}{cccccc}
		\hline
		k	& a & b  & $\epsilon_0$ & $\mu_1$ & $\mu_2$  \\
		\hline
		8.0	& 0.01   & 0.15      & 0.002 & 0.2 & 0.3  \\
		\hline	
	\end{tabular}
	\label{tab:ap-2}
\end{table}
The diffusion coefficient for the myocardium are set as  
$d^M_{iso} = 0.8\text{mm}^2 / \text{ms}$ and $d^M_{ani} = 1.2 \text{mm}^2 / \text{ms}$, 
and for the Purkinje network $d^P_{iso} = 22 \text{mm}^2 / \text{ms}$,
indicating that the conduction through the network is $27.5$ times faster than that in the myocardium \cite{costabal2016generating}. 
\subsubsection{Electrophysiology}\label{sec:lv-electrophysiology}
In this part, 
we present the numerical results of the transmembrane potential propagates 
in the left ventricle with inclusion of the Purkinje network with different excitation strategies. 

\begin{figure}[htb!]
	\centering
	\includegraphics[trim = 1mm 2mm 1mm 1mm, clip, width=0.95\textwidth]{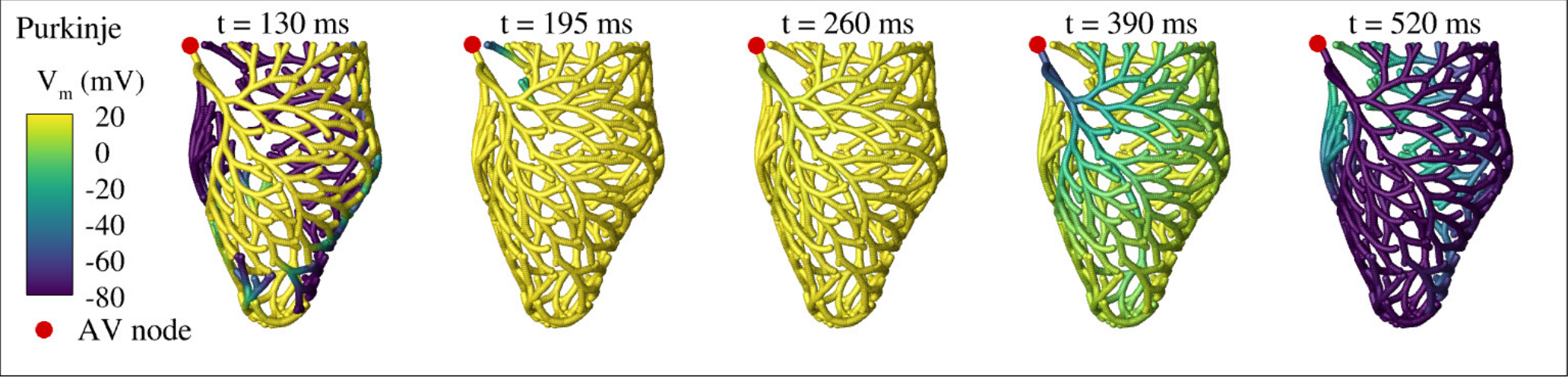}
	\includegraphics[trim = 1mm 2mm 1mm 1mm, clip, width=0.95\textwidth]{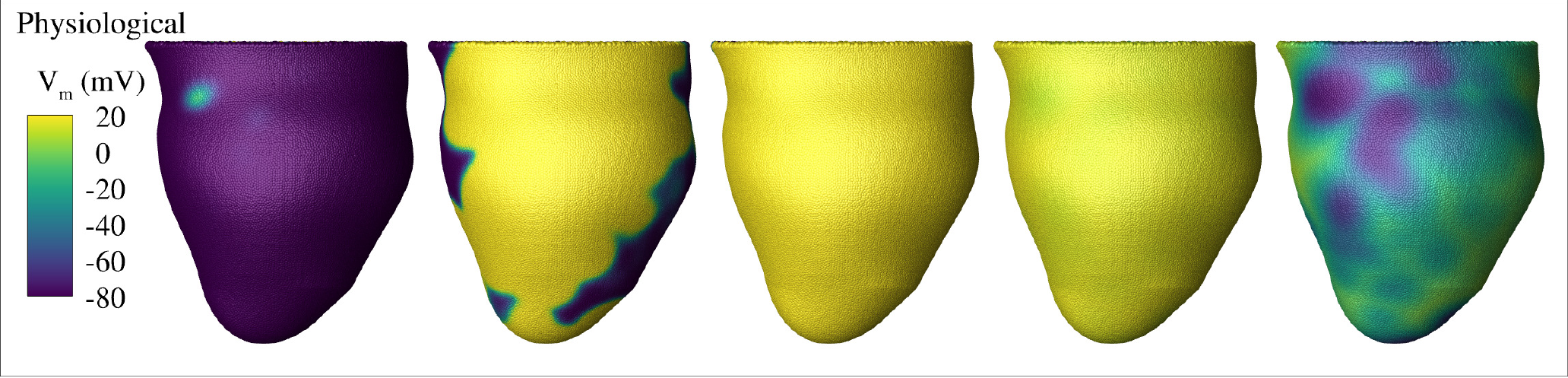}
	\includegraphics[trim = 1mm 2mm 1mm 1mm, clip, width=0.95\textwidth]{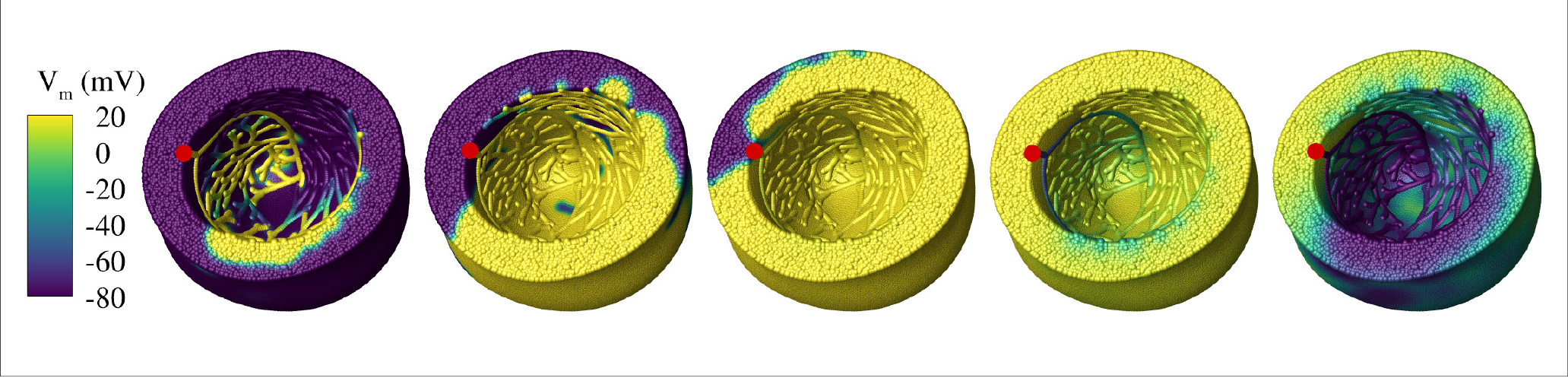}
	\includegraphics[trim = 1mm 2mm 1mm 1mm, clip, width=0.95\textwidth]{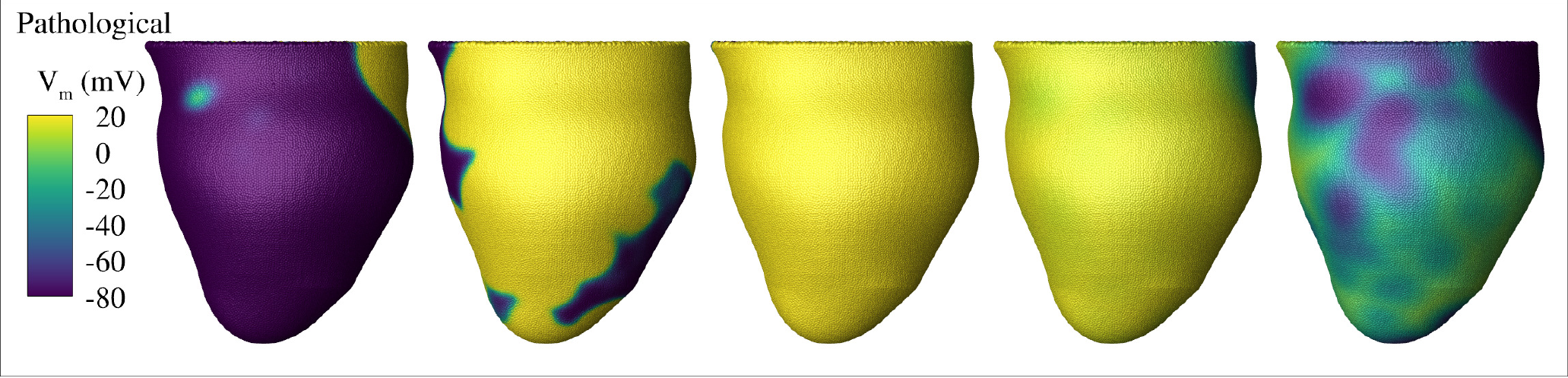}
	\includegraphics[trim = 1mm 2mm 1mm 1mm, clip, width=0.95\textwidth]{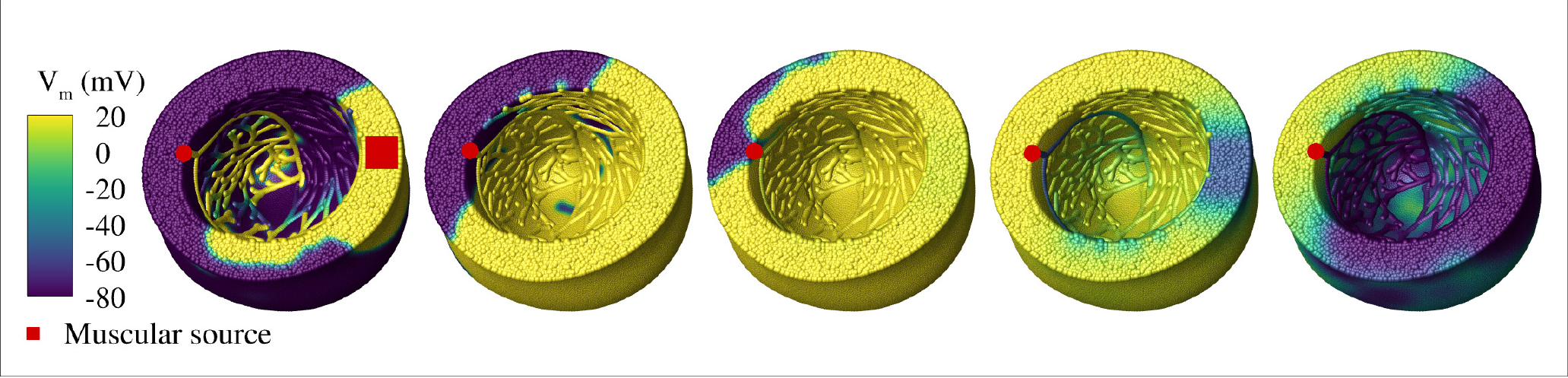}
	\includegraphics[trim = 1mm 2mm 1mm 1mm, clip, width=0.95\textwidth]{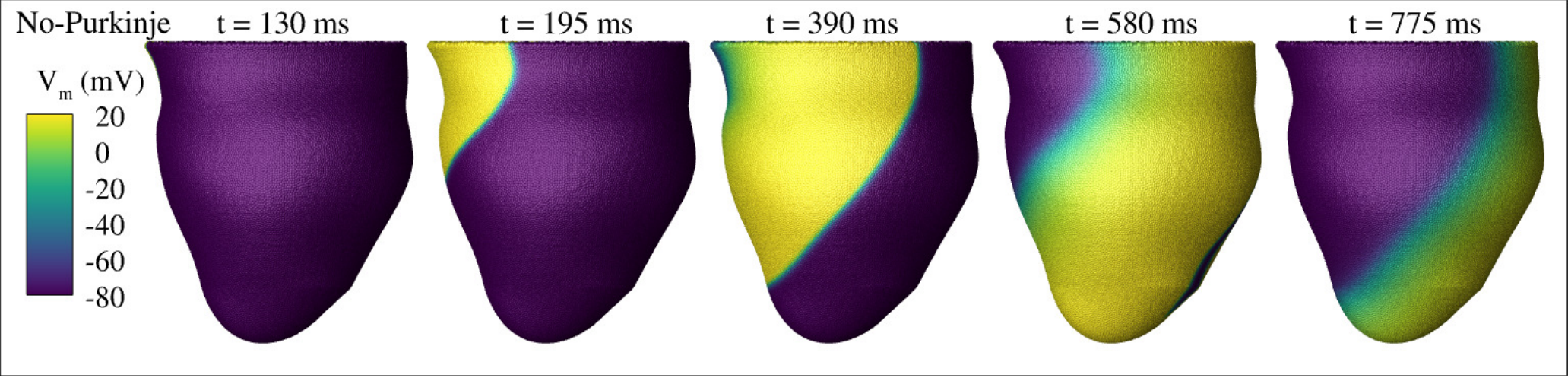}
	\includegraphics[trim = 1mm 2mm 1mm 1mm, clip, width=0.95\textwidth]{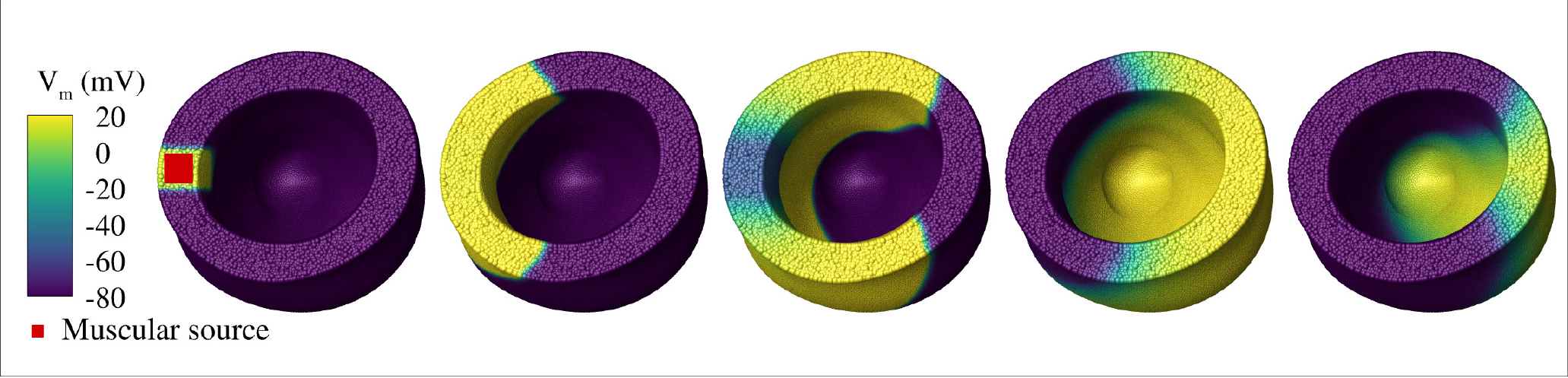}
	\caption{Left ventricle with the Purkinje network: 
		Activation consequences of the ventricle with the Purkinje network under physiological healthy and pathological conditions, 
		and without the network. 
		(For interpretation of the references to color in this figure legend, the reader is referred to the web version of this article.)
	}
	\label{figs:pkj-lv}
\end{figure}

Figure \ref{figs:pkj-lv} reports the transmembrane potential at different time instants through the Purkinje network 
and in the left ventricle with inclusion of the Purkinje network under physiological and pathological conditions, 
and without the network. 
In the Purkinje network, 
the transmembrane potential activated by a stimulus of $V_m = 1.0$ 
initiated at the AV node rapidly travels through the network 
in the free-pulse pattern due to its faster conduction velocity. 
At the PKJs, 
which is represented by the terminal particles, 
the potential enters the myocardium and excites the apex of the left ventricle under physiological healthy condition, 
allowing non-smooth propagation of wave front. 
Under pathological condition, 
which is characterized by an extra muscular source located in the opposite region with respect to the AV node, 
the left ventricle is excited similarly to the healthy condition with more complex wave front collisions are noted. 
Note that the pathological test demonstrates the suitability of the present method 
in view of the solutions of the MM coupled problem when complex fronts propagate. 
Without the Purkinje network, 
the left ventricle is activated via slow smooth diffusion of the transmembrane potential 
starting at the muscular source, 
which is located in the region of AV node, 
and propagating from the base to the apex. 
Compared with the results obtained without the Purkinje network, 
the activation sequences with the inclusion of the network 
exhibits non-smooth potential propagation and rapid excitation from the apex to the base. 

\begin{figure}[htb!]
	\centering
	\includegraphics[trim = 1mm 1mm 1mm 1mm, clip, width=\textwidth]{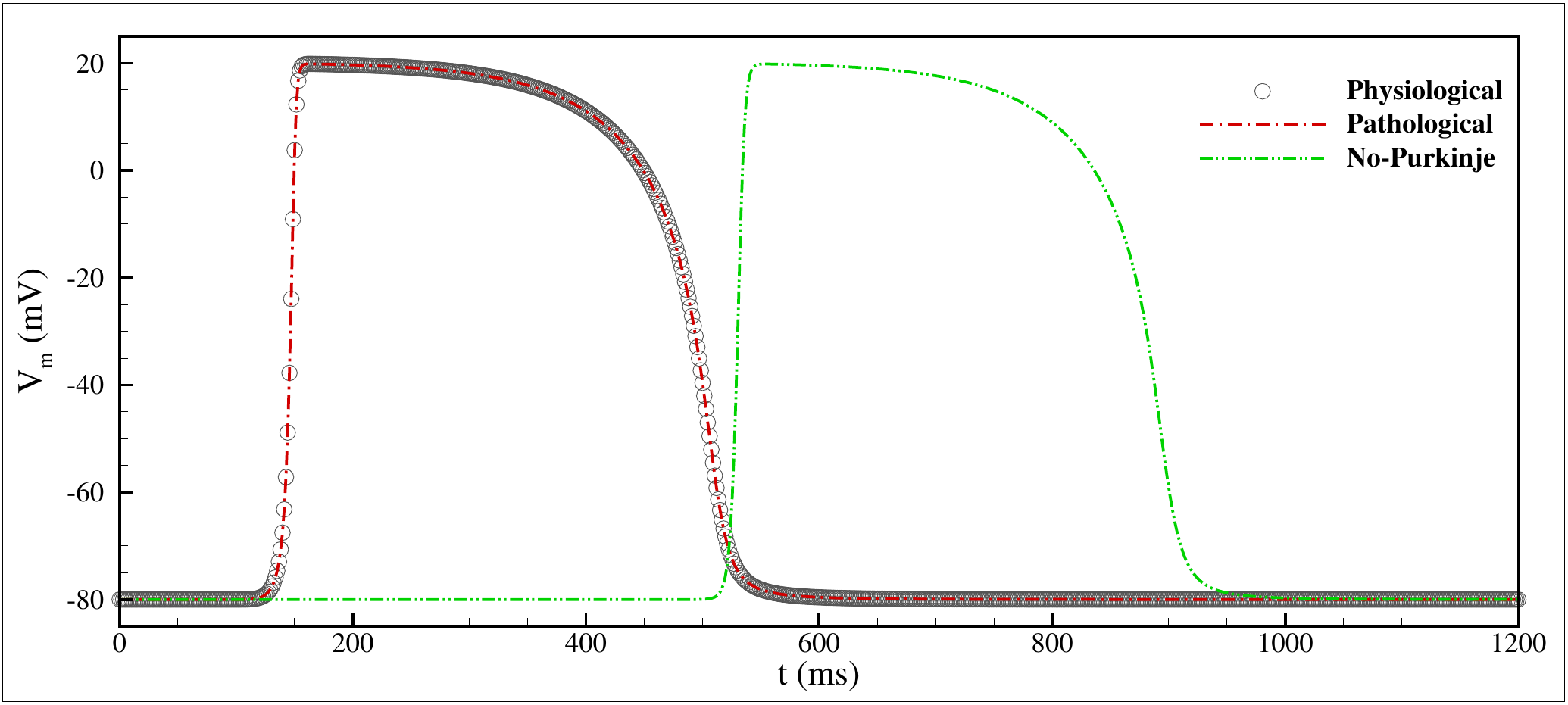}
	\caption{Left ventricle with the Purkinje network: 
		The evolution of transmembrane potential $V_m$ recorded at the apex of the ventricle with the Purkinje network 
		under physiological and pathological conditions, 
		and without the network. 
		(For interpretation of the references to color in this figure legend, the reader is referred to the web version of this article.)
	}
	\label{figs:pkj-lv-data}
\end{figure}

Figure \ref{figs:pkj-lv-data} illustrates the time evolutions of the transmembrane potential at the apex with 
the Purkinje network under physiological and pathological conditions,  
and without the network. 
With the purkinje network, 
the potential propagation under physiological and pathological conditions present identical profile indicating 
that the excitation in the apex is dominated by the signals from the network. 
Compared with the results obtained without the network, 
the apex is activated much more rapid as expected under 
both physiology and pathological conditions.  

The results reported herein show that the present multi-order SPH method could be applied successfully to model the transmembrane potential 
propagates in realistic left ventricle with inclusion of the Purkinje network. 
This represents a crucial step in view of solving a complete electromechanics coupling problem
of a realistic ventricle in presence of the Purkinje network. 
%
\subsubsection{Electromechanics}\label{sec:lv-electromechanics}
In this part, 
we assess the versatility of the present method for modeling the excitation-contraction 
of the realistic left ventricle with inclusion of the Purkinje network. 
Following the previous study for electrophysiology, 
we consider the numerical approximation of the active electromechanical response in the left ventricle with different excitation strategies. 
For the mechanical response, 
we apply Holzapfel-Ogden model \cite{holzapfel2009constitutive} 
with the constant parameters given in the Table \ref{tab:ho-model1} 
and the active cardiomyocite contraction stress $T_a = 0.1 ~\text{kPa}$. 
For simplicity, 
the displacement degrees of freedom on the top base of the left ventricle 
are constrained and the whole heart surface is assumed to be flux-free. 
\begin{table}[htb!]
	\centering
	\caption{Left ventricle with the Purkinje network: Parameters for the Holzapfel-Ogden constitution model \cite{holzapfel2009constitutive}.}
	\begin{tabular}{cccc}
		\hline
		$a = 0.059$ kPa	& $a_f = 18.472$ kPa & $a_s = 2.841$ kPa  & $a_{fs} = 0.216$ kPa  \\
		\hline
		$b = 8.023$ 	& $b_f = 16.026$    & $b_s = 11.12$      & $b_{fs} = 11.436$   \\
		\hline	
	\end{tabular}
	\label{tab:ho-model1}
\end{table}

\begin{figure}[htb!]
	\centering
	\includegraphics[trim = 1mm 1mm 8mm 1mm, clip, width= \textwidth]{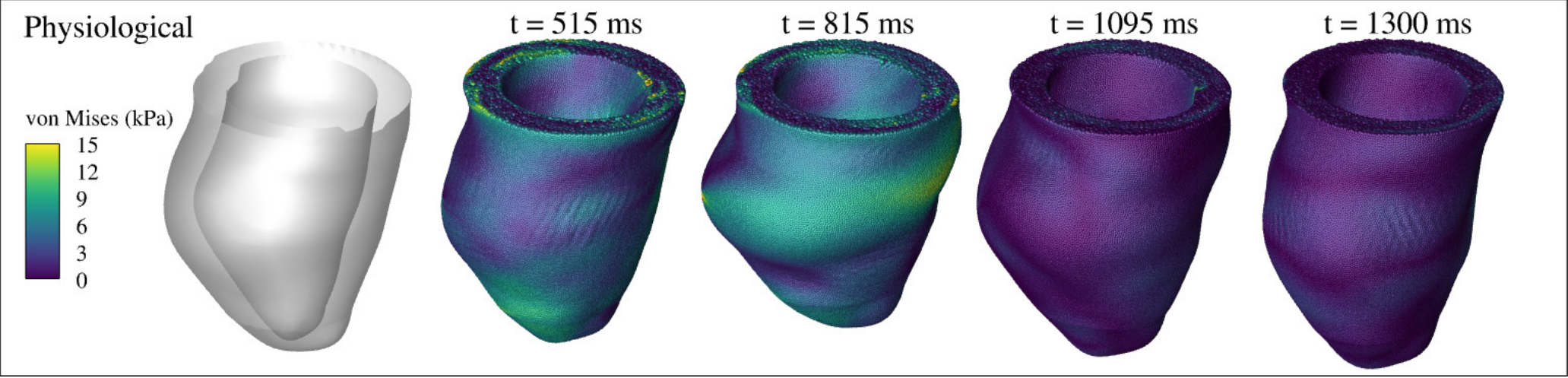}
	\includegraphics[trim = 1mm 1mm 8mm 1mm, clip, width= \textwidth]{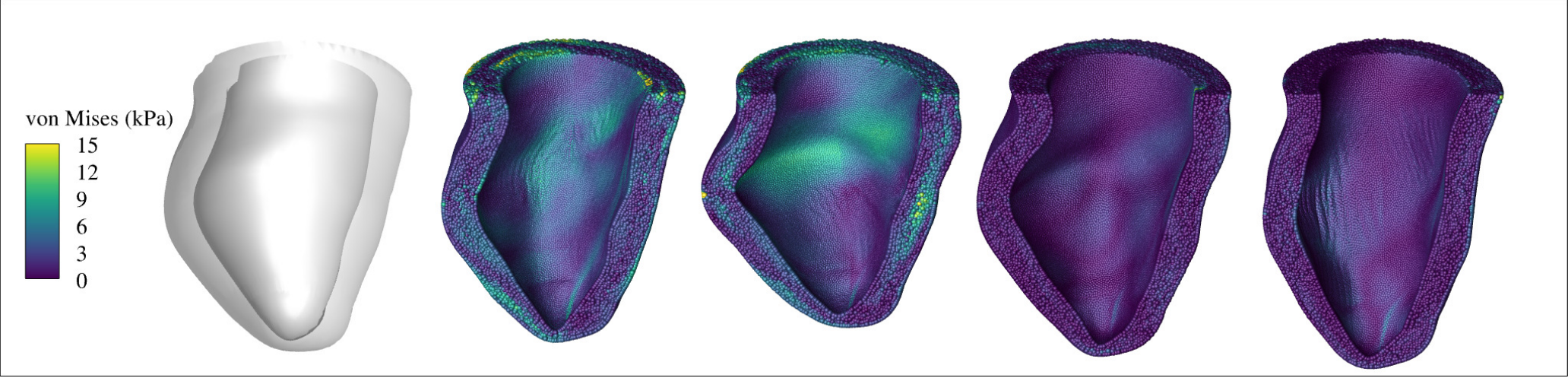}
	\includegraphics[trim = 1mm 1mm 8mm 1mm, clip, width= \textwidth]{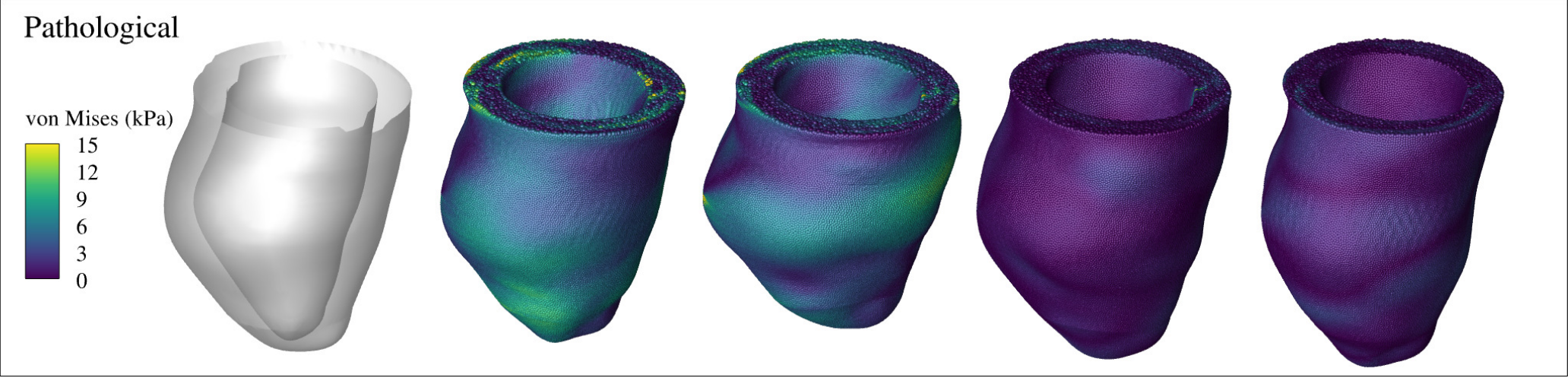}
	\includegraphics[trim = 1mm 1mm 8mm 1mm, clip, width= \textwidth]{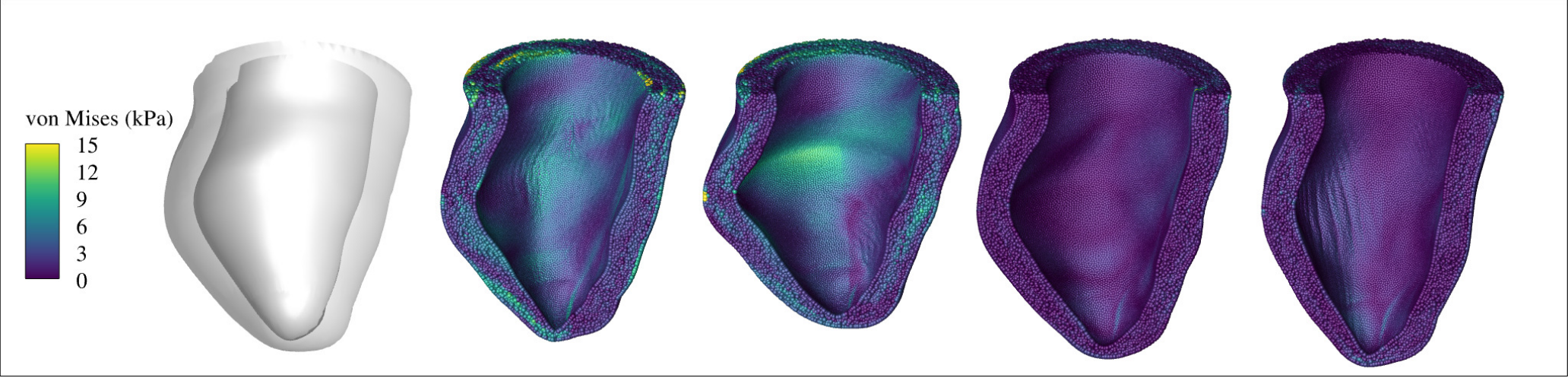}
	\includegraphics[trim = 1mm 1mm 8mm 1mm, clip, width= \textwidth]{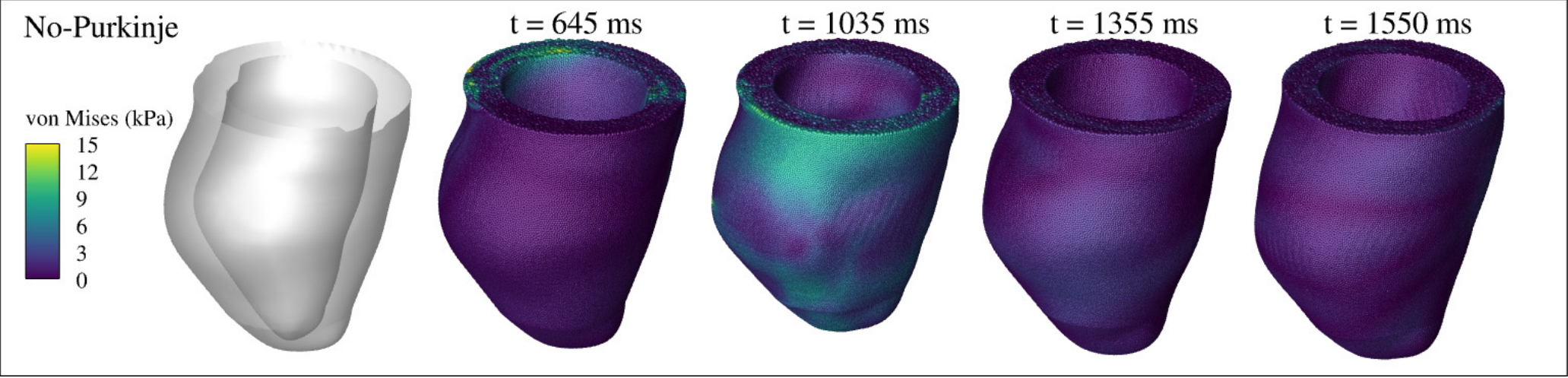}
	\includegraphics[trim = 1mm 1mm 8mm 1mm, clip, width= \textwidth]{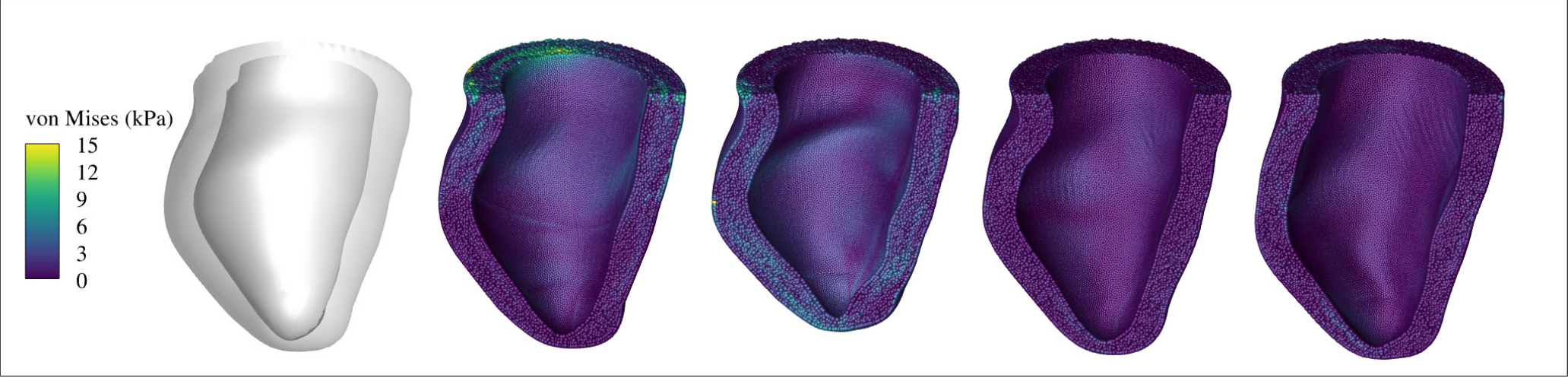}
	\caption{Left ventricle with the Purkinje network: 
		Activation-contraction consequences of the ventricle with the Purkinje network under physiological and pathological conditions, 
		and without the network. 
		For comparison, 
		the initial geometry representation is presented in the left panel. 
		(For interpretation of the references to color in this figure legend, the reader is referred to the web version of this article.)
	}
	\label{figs:pkj-lv-contraction}
\end{figure}

Figure \ref{figs:pkj-lv-contraction} reports excitation-contraction configurations 
with von Mises stress contours of the left ventricles with the Purkinje network under 
physiological and pathological conditions, 
and without the network. 
For rigorous comparison, 
the undeformed configuration of the left ventricle is also presented in Figure \ref{figs:pkj-lv-contraction} (left panel).
In general, 
similar contraction pattern, 
i.e.,
the excitation-contraction gives rise to the up- and down-ward motions of the apex as 
the depolarization front traveling in the ventricles, 
are observed in three conditions. 
Also, 
the apex's down-ward motion is accompanied by 
the physiologically observed wall thickening and the overall torsional motion of the ventricle. 
Due to the inhomogeneous myocyte orientation distribution incorporated with the anisotropic material model, 
the physiologically active response through the non-uniform contraction of myofibers are noted. 
However, 
compared with the results obtained without the Purkinje network, 
some remarkable differences can be noted in the deformation pattern due to 
different electrical activation profiles introduced by the inclusion of the Purkinje network. 
For example,
notable large von Mises stress is exhibited in the regions close to the apex due the fact that 
the excitation with the inclusion of the Purkinje network is propagates from the apex to the base. 
Also, 
the left ventricle shows larger deformation and more rapid excitation when the Purkinje network is included. 
As for the physiological and pathological excitations, 
similarly contraction pattern is noted with small difference in excitation time which will be shown in the following section. 

\begin{figure}[htb!]
	\centering
	\includegraphics[trim = 2mm 2mm 2mm 2mm, clip, width=\textwidth]{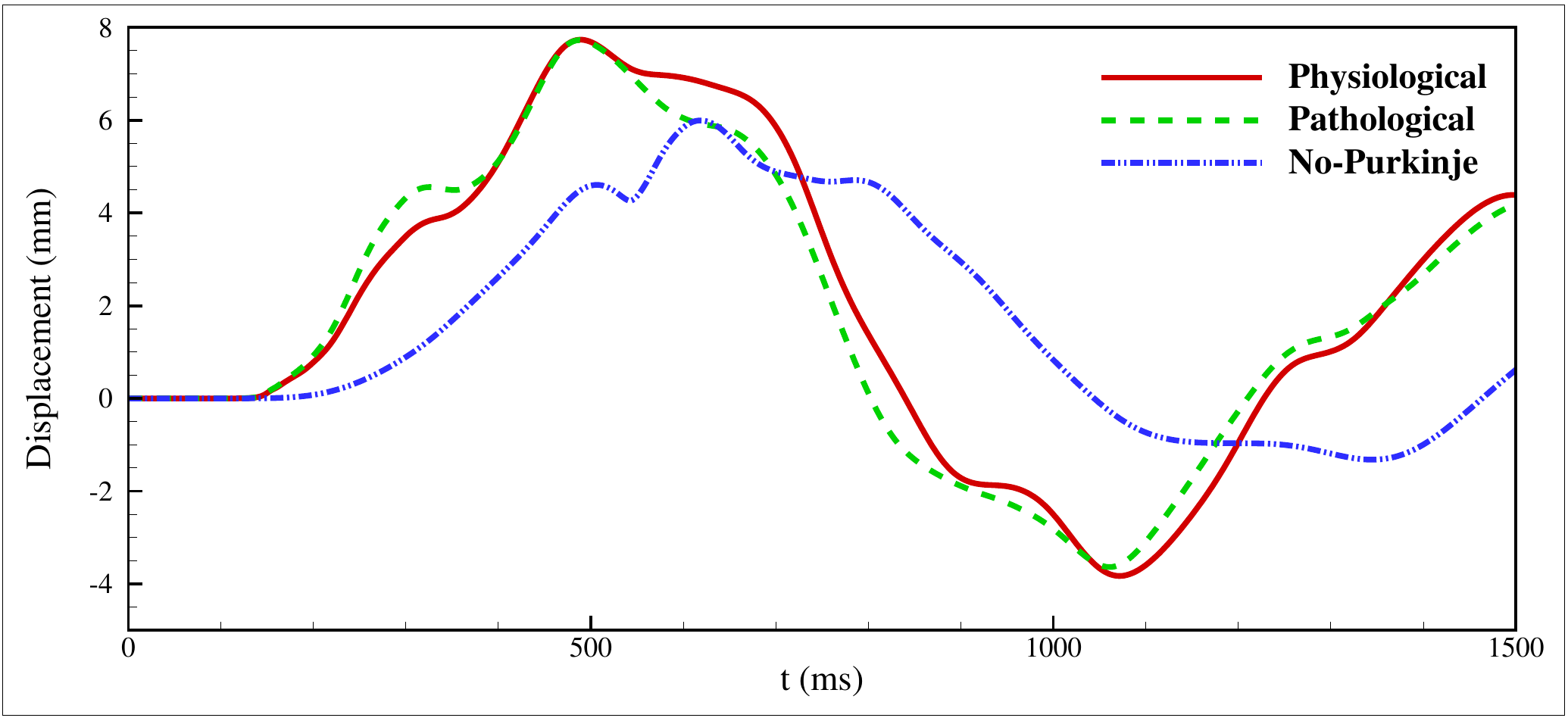}
	\caption{Left ventricle with the Purkinje network: 
		The time history of displacement in the $z$-axis of the apex of the ventricle with the Purkinje network 
		under physiological and pathological conditions, 
		and without the network. 
		(For interpretation of the references to color in this figure legend, the reader is referred to the web version of this article.)
	}
	\label{figs:pkj-lv-contraction-data}
\end{figure}

Figure \ref{figs:pkj-lv-contraction-data} 
illustrates the time evolution of the $z$ components of the displacements at the apex with three excitation strategies. 
With inclusion of the Purkinje network, 
the left ventricle shows larger displacement of the apex with rapid excitation. 
Compared with the physiological condition, 
the pathological one induces slightly quick excitation-contraction mechanical response 
in the left ventricle due to the contraction activated by the muscular current source.  

The results reported herein show that the present method could be applied successfully 
to solve the electromechanics coupling problem
of the realistic left ventricle with inclusion of the Purkinje network. 
This represents a crucial step in view of investigating the 
excitation-contraction profile of an anatomical high-resolution heart model in presence of the Purkinje network, 
which will be the main objective of our future work. 
%
\subsubsection{Computational efficiency}\label{sec:computationalefficiency}
In this paper, 
we have proposed a multi-time stepping scheme for the time integration of the active electromechanical response of the myocardium 
with inclusion of the Purkinje network. 
This part is devoted to rigorously assess the computational performance of the proposed multi-time stepping scheme. 
We analyze the total CPU time for the simulation of the excitation-contraction of the left ventricle with the presence of the 
Purkinje network and the computer information is given in Section \ref{sec:efficiency}. 
Note that the single-time stepping denotes the time integration where the minimal value of 
$\Delta t_d^P$,  $\Delta t_m^M$ and  $\Delta t_m^M$ is chosen as the time step size.  

Table \ref{tab:pkj-lv-cpu-time} reports the computation time with different time stepping schemes 
for the electrophysiology and electroemchanics problems of the left ventricle with inclusion of the Purkinje network.
It can also be observed that with the present multi-time stepping scheme,  
one is able to obtain speedup of $36.6$ and $28.7$ for electrophysiology and electroemchanics simulations, respectively. 
\begin{table}[htb!]
	\centering
	\caption{Left ventricle with the Purkinje network: Computational efficiency. 
		Here, we evaluate the CPU wall-clock time for computation until the physical time of $10 \text{ms}$. }
	\begin{tabular}{cccc}
		\hline
		Cases      			& 	Multi-time stepping & Single-time stepping  & Speedup \\
		\hline
		Electrophysiology	&   $5.59 \text{s}$ &  $204.61 \text{s}$  & $36.6$ \\ 
		\hline
		Electromechanics	&   $11.79 \text{s}$   & $338.52 \text{s}$  &  $28.7$ \\ 
		\hline
	\end{tabular}
	\label{tab:pkj-lv-cpu-time}
\end{table}
%
%
%
\section{Concluding remarks}\label{sec:conclusion}
As a pioneering work, 
this paper presents a multi-order SPH method for cardiac electrophysiology and electromechanics with 
inclusion of the Purkinje network. 
The main novelties of the present work are summarized as follows. 
To the best knowledge of the authors, 
the following aspects were addressed herein for the first time in developing a meshless approach for total heart modeling, 
\begin{enumerate}
	\item We introduced an efficient algorithm by exploiting level-set geometry presentation 
				and CLL algorithm for network generation on arbitrarily complex surface. 
	\item We proposed a reduced-order SPH method to resolve the electrical activation in the Purkinje network 
				by solving one-dimensional monodomain equation.
	\item We developed a multi-order coupling paradigm to capture the coupled nature of propagation 
				arising from the interaction between the Purkinje network and the myocardium. 
	\item We presented a multi-time stepping algorithm to optimize the computational efficiency for modeling the electromechanics 
	coupling problem of the myocardium with inclusion of the Purkinje network. 
\end{enumerate}
Ultimately, 
comprehensive and rigorous studies of 
the potential propagation in myocardium fiber, 
cubiod myocardium with inclusion of a generic network, 
electrophysiology and electromechanics coupling problems in the left ventricle with inclusion of the Purkinje network 
have been conducted. 
Also, 
the CPU time is analyzed to assess the computational efficiency of the present network generation and the multi-time stepping algorithms. 
The results demonstrate the robustness, accuracy and feasibility of the proposed SPH method 
for cardiac electrophysiology and electromechanics in realistic ventricle with the presence of the Purkinje network. 

The multi-order SPH methods developed in this work is an essential component of an unified meshless approach 
to accurately describing the electrical activation in the left and right ventricles. 
Base on this work and the previous one \cite{zhang2021integrative},  
the long-term objective of this serial study is developing a multi-physics total-function heart simulator, 
which has the potential to complement and extend human understanding of cardiac diseases.
In particular,
one import improvement on the network-myocardium coupling problem would be taking 
the “pull and push” effect \cite{kucera1998slow, kucera2001mechanistic, vergara2016coupled}, 
which is due to the fact that the current just before the bifurcation point needs 
to increase its value in order to be able to stimulate the increased number of cells after the bifurcation,
into consideration. 
Ultimately,  
we would like to move to the coupled electromechanics problem of 
a high-resolution anatomical heart model with four chambers 
to explore the influence of the patient-specific Purkinje network \cite{palamara2015effective} 
on local tissue strain distributions and global pressure-volume loops. 
These insights will help the research community elucidate the interplay between 
electrical conduction disturbances and the loss of mechanical function. 
%
%
\clearpage
\section*{CRediT authorship contribution statement}
{\bfseries  Chi Zhang:} Conceptualization, Methodology, Investigation, Visualization, Validation, Formal analysis, Writing - original draft, Writing - review \& editing; 
{\bfseries  Hao Gao:} Investigation, Writing - review \& editing;
{\bfseries  Xiangyu Hu:} Supervision, Methodology, Investigation, Writing - review \& editing.
%
%
\section*{Declaration of competing interest }
The authors declare that they have no known competing financial interests 
or personal relationships that could have appeared to influence the work reported in this paper.
%
%
\section*{Acknowledgement}
C. Zhang and X.Y. Hu would like to express their gratitude to Deutsche Forschungsgemeinschaft (DFG) 
for their sponsorship of this research under grant numbers DFG HU1527/10-1 and HU1527/12-1.
%
%
\section*{References}
\bibliography{mybibfile}
%
%
\end{document}